\documentclass[conference]{IEEEtran} 

\IEEEoverridecommandlockouts
\usepackage{amsmath,amssymb}
\usepackage{mathtools}
\usepackage{graphicx,epsfig}
\usepackage{setspace} 
\usepackage{svg}
\usepackage{multirow}
\usepackage{bm}
\usepackage{mwe}
\usepackage{amsmath}
\usepackage{caption}
\usepackage{subcaption}
\usepackage{placeins}
\usepackage[hyphens]{url}
\usepackage[hidelinks]{hyperref}
\hypersetup{breaklinks=true}
\usepackage{cite}
\usepackage{soul}
\usepackage{bbm}
\usepackage{graphicx}
\usepackage{lettrine}
\usepackage{verbatim} 
\usepackage{calrsfs}
\DeclareMathAlphabet{\pazocal}{OMS}{zplm}{m}{n}

\usepackage{pifont}
\newcommand{\xmark}{\ding{55}}%

\usepackage[english]{babel}
\usepackage[utf8]{inputenc}
\usepackage{algorithm}
\usepackage[noend]{algpseudocode}
\algdef{SE}[DOWHILE]{Do}{doWhile}{\algorithmicdo}[1]{\algorithmicwhile\ #1}%

\usepackage{array}
\newcolumntype{P}[1]{>{\centering\arraybackslash}p{#1}}

\def\myplotScale{1}
\def\myFigScale{1}

\begin{document}
\captionsetup[figure]{labelformat={default},labelsep=period,name={Fig.},font=footnotesize}  
\title{5G-Advanced AI/ML Beam Management: Performance Evaluation with Integrated ML Models}
\author{\IEEEauthorblockN{Nalin Jayaweera$^{1,2}$, Andrea Bonfante$^{1}$, Mark Schamberger$^{1}$, Amir Mehdi Ahmadian Tehrani$^{1}$, \\Tachporn Sanguanpuak$^{1}$, Preetish Tilak$^{1}$, Keeth Jayasinghe$^{1}$, Frederick W. Vook$^{1}$, Nandana Rajatheva$^{2}$}
		\IEEEauthorblockA{$^{1}$Nokia Standards, $^{2}$University of Oulu}
	}

\maketitle

\begin{abstract}
The legacy beam management (BM) procedure in 5G introduces higher measurement and reporting overheads for larger beam codebooks resulting in higher power consumption of user equipment (UEs). Hence, the 3rd generation partnership project (3GPP) studied the use of artificial intelligence (AI) and machine learning (ML) in the air interface to reduce the overhead associated with the legacy BM procedure. The usage of AI/ML in BM is mainly discussed with regard to spatial-domain beam prediction (SBP) and time-domain beam prediction (TBP). In this study, we discuss different sub-use cases of SBP and TBP and evaluate the beam prediction accuracy of AI/ML models designed for each sub-use case along with AI/ML model generalization aspects. Moreover, a comprehensive system-level performance evaluation is presented in terms of user throughput with integrated AI/ML models to a 3GPP-compliant system-level simulator. Based on user throughput evaluations, we present AI/ML BM design guidelines for the deployment of lightweight, low-complexity AI/ML models discussed in this study. 
\end{abstract}


\section{Introduction} 
\IEEEpubidadjcol

\IEEEPARstart{F}{ive} G-Advanced (5G-Advanced) wireless networks herald the prospect of a thoroughly interconnected and mobile society characterized by the expectation of high throughput, energy efficient, immersive and data-driven radio access networks (RANs) \cite{NokiaWhitepaperRel18, NokiaWhitepaperRel19}. Coupled with data-driven approaches, RAN has the potential to develop predictive capabilities that learn from the environment through the use of artificial intelligence (AI) and machine learning (ML) techniques \cite{8755300}. 
Therefore, many companies are presently integrating AI/ML technologies into the existing 5G network and devices, with AI/ML emerging as a pivotal technology cornerstone for 5G-Advanced once standardized within the air interface \cite{9569393, 9605055, 10017176}. AI/ML 5G-Advanced can assist the transition to the next generation of networking, 6G, which is envisioned to feature an even more pronounced usage of AI/ML technologies in mobile communications as inherently designed based on AI-native interface \cite{9446676, 10156818, 9768336}. 

The recently concluded 3rd generation partnership project (3GPP) Rel-18, has studied the support of a new AI/ML-enabled radio interface for the next cellular systems \cite{3gpp38843}. 
Beam management (BM) represents one of the different pilot use cases for applying the AI/ML model in the new radio (NR) air interface. These pilot use cases aim to identify a common AI/ML framework, including functional requirements of AI/ML architecture, which could be used in subsequent releases and will pave the way for the design of native AI/ML 6G networks. The motivation behind the study of this use case stems from the increasing demand for efficient beam prediction schemes. One of the primary objectives of the 
 3GPP Rel-18 study item (SI) was to investigate the application of AI/ML-driven BM schemes for reducing the measurement overhead -- providing more flexibility for beam measurements and a larger room for the user equipment (UE) power saving -- while maintaining comparable end-to-end performance of legacy BM procedures. 

In the legacy BM procedure, the gNodeB (gNB) configures UEs to measure reference signal (RSs) resources including synchronization signal block (SSB) and channel state information RSs (CSI-RS). UEs measure and provide feedback to a gNB to assist the legacy beam alignment process performed with P1, P2 and P3 procedures \cite{5GNRbook}. In the initial beam-pair establishment procedure (P1), gNB transmits SSB resources employing a beam codebook referred to as SSB beams, and the UE measures the reference signal received power (RSRP) of each SSB beam with receiver (RX) beams and reports the strongest set of RS identifiers and corresponding RSRP measurements. The beam refinement procedure (P2) finds the UE-specific CSI-RS beams by transmitting CSI-RS beams, which are more directional and provide higher received signal strength than SSB beams. Hence, CSI-RS beams are used for downlink (DL) data transmission. In the receiver beam refinement procedure (P3), gNB transmits the same transmitter (TX) beam over different time instances, and UE measures determine the best RX beam. 
The large number of beams included in TX and RX codebooks leads to performing a large number of measurements for determining the best TX-RX beams. 

The legacy BM process is time-inefficient and not scalable when the size of antenna arrays increases. ML algorithms can replace sequential beam sweeping by predicting beams in both time and spatial domains as proposed in 3GPP Rel-18 BM use cases. The scope of spatial-domain beam prediction (BM-Case1) is to predict the best DL TX beam or DL TX/RX beam pairs in different spatial locations. Conversely, time-domain beam predictions (BM-Case2) aim to predict the best DL TX beam or DL TX/RX beam pairs to use for future time instances. It has been shown that ML algorithms enable prediction of the serving beam for different spatial locations and time instances, thereby avoiding the necessity for exhaustive beam measurements \cite{8734054, 9898910, 9512417}. 

\subsection{Relevant Work}

The application of AI/ML for BM has attracted many research efforts in recent years. In \cite{Li2020}, the legacy BM procedures and associated challenges are explained while presenting future trends such as AI/ML-based BM to overcome those challenges. Additionally, many studies and findings account for AI/ML BM in the literature. In \cite{lin2023overview}, an overview is presented on the 3GPP Rel-18 studies on AI/ML to improve the 5G NR air interface and discuss the AI/ML BM framework and its functionality. Similarly, in-depth discussions related to AI/ML-based BM procedures and the standardization aspects are presented in \cite{Xue2023}.

Many studies account for spatial-domain and time-domain beam prediction in the literature \cite{10412253}. \cite{Zuo2022} investigates an ML-based BM scheme with less beam measurement overhead. The authors present a convolutional neural network (CNN) based ML model for beam pair prediction for a fixed beam measurement configuration, and another ML model to address the generalization aspect. Next, transfer learning-based enhancement is proposed to reduce ML model training time and data requirements. However, this work explores the ML-based spatial-domain beam predictions and evaluation is limited for beam prediction accuracy-related key performance indicators (KPIs). AI/ML-based time-domain prediction algorithms are presented in \cite{10123939} with beam prediction accuracy and system evaluations. Additionally, the proposed models are integrated into a 3GPP-compliant gNB and a UE to evaluate the model performance in a test setup. The study \cite{10334007} performs time-domain beam prediction and evaluates beam prediction accuracy-related KPIs to evaluate important aspects such as UE speed and required historical data volume to make better decisions. Wang et al.\cite{10419651} study spatial-domain beam prediction and time-domain beam pair prediction, and show AI-based algorithms outperform non-AI/ML based solutions using beam pair prediction accuracy. Among all AI/ML-based beam tracking methods, \cite{10200730} proposes a Bayesian optimization-based best beam tracking technique that requires no or little training on historical data. Furthermore, authors in \cite{10335766} propose a AI/ML model activation and deactivation framework to reduce unnecessary ML model switching by $10\%$.

In addition to the RS measurement-based beam predictions, assisted information such as location data is used to improve AI/ML prediction accuracy. A recurrent neural network (RNN) based beam-pair prediction algorithm is proposed in \cite{9417509} which used beam measurements as well as sensor data to enhance the prediction accuracy. Light detection and ranging (LiDAR) sensor information assisted time-domain beam prediction is proposed in \cite{10012751} that reduced RS overhead by $13.2\%$ for a best beam prediction accuracy of $88.7\%$. 

Most of the available studies explore spatial-domain or time-domain beam prediction and evaluation is mainly performed with beam prediction accuracy-related KPIs. A few works have considered system-level performance evaluation. System-level simulation (SLS) studies based on 3GPP NR Rel-16 were performed in \cite{9562975}. More recently, the work proposed in \cite{10123939}, focused on SLS evaluation of time-domain beam prediction adopting the latest UE mobility and orientation model considered in 3GPP 5G-Advanced NR Rel-18. 

\subsection{Contributions}
This paper presents a comprehensive analysis considering system-level performance evaluation methodologies for spatial-domain beam prediction and time-domain beam prediction by integrating the ML models studied in 3GPP Rel-18 study item (SI) into a 3GPP-based system-level simulator. The work considered in \cite{10123939} has primarily focused on time-domain scenarios, overlooking the spatial-domain BM use case that may be more commonly applied in practical environments, where most of the UEs are stationary. As a result, there is a growing need for more comprehensive studies that show the effectiveness of ML techniques to accurately predict beams both in the spatial-domain and time-domain. This paper addresses this need by proposing novel results that consider both spatial-domain and time-domain beam prediction models. We integrate these schemes into a state-of-the-art system-level simulator and we evaluate them using the latest 5G-Advanced
Rel-18 assumptions. The performance evaluation is not limited to the beam prediction accuracy, as compared to most of the existing prior works, but considers system-level KPIs such as user throughput and measurement overhead which shows the realistic gains of using AI/ML for BM. The primary aim is to provide new insights about the link between ML model performance and system-level performance. The findings are summarized into design guidelines to select suitable AI/ML models for UEs with different channel conditions. Finally, we highlighted future research directions based on the observations of this study. 

The rest of the paper is organized as follows. In Section II, we discuss the use cases of AI/ML-based BM with problem formulations. AI/ML BM procedures are presented in Section III while the simulation framework is explained in Section IV with ML model descriptions and beam prediction KPIs. In Section V, we present an in-depth analysis of the simulation results including beam prediction accuracy and system level performance. Then we present AI/ML BM design guidelines based on the evaluation results. Finally, we conclude the paper in Section VI with the main insights of this study.       

\section{AI/ML Based BM Use-Cases} 
We consider a network deployment of gNBs each one serving multiple UEs according to an NR system operating in FR2 (frequency range 2). Both gNBs and UEs use analog beamforming with a single radio frequency (RF) chain connected to multiple antenna elements to form beams in spatial directions according to the beamforming vectors $\textbf{b}_{x}$ where $x \in \{tx,rx\}$ is a notation to indicate the transmitter and receiver \cite{AnalogBF}. The gNBs and UEs are equipped with uniform planar arrays (UPAs) having antenna spacing $\lambda/2$ and formed by a number of antennas $M_{x} = M^H_{x} \times M^V_{x}$ with spacing $d_H$, $d_V$ in the horizontal and vertical directions where $M_x^V$ and $M_x^H$ denote the vertical and the horizontal dimension of the UPAs, respectively. 
Additionally, $\theta_{tx}$ and $\phi_{tx}$ denote the azimuth and elevation angle of departure (AoD), and similarly $\theta_{rx}$ and $\phi_{rx}$ denote the azimuth and elevation angle of arrival (AoA). The TX beamforming vector $\textbf{b}_{tx}(\theta_{tx}, \phi_{tx})$ or RX beamforming vector $\textbf{b}_{rx}(\theta_{rx}, \phi_{rx})$ used to steer the beam is generalized and expressed as $\textbf{b}_{x}(\theta_{x}, \phi_{x})$ that can be defined as following $M_x^H \times M_x^V$ matrix:
\begin{equation}
\footnotesize
\begin{split}
     &\frac{1}{\sqrt{M_x^H}} [1,e^{-j2 \pi \frac{d_H}{\lambda}\text{sin}(\phi_x)\text{cos}(\theta_x)},..., e^{-j2 \pi \frac{d_H}{\lambda} (M_x^H-1)\text{sin}(\phi_x)\text{cos}(\theta_x)}]^\top \otimes\\
    &\frac{1}{\sqrt{M_x^V}} [1,e^{-j2 \pi \frac{d_V}{\lambda}\text{cos}(\phi_x)},..., e^{-j2 \pi \frac{d_V}{\lambda}(M_x^V-1)\text{cos}(\phi_x)}],
\end{split}
\end{equation}
where $j = \sqrt{-1}$, $\otimes$ denotes the Kronecker product and $[.]^\top$ performs the transpose operation.

Next, we consider a vector of TX beam codebook composed of $N_{tx} = N_{tx}^A \times N_{tx}^E$ beams, which can be defined as 
\begin{equation}
\begin{split}
   \text{CB}_{tx} = &\{ \textbf{b}_{tx}(\theta_{tx}^1, \phi_{tx}^1),  \textbf{b}_{tx}(\theta_{tx}^2, \phi_{tx}^1),...,  \textbf{b}_{tx}(\theta_{tx}^{N^A_{tx}}, \phi_{tx}^1), \\
   & \textbf{b}_{tx}(\theta_{tx}^1, \phi_{tx}^2),...,  \textbf{b}_{tx}(\theta_{tx}^{N_{tx}^A}, \phi_{tx}^{N_{tx}^E}) \},
\end{split}
\end{equation}
where subscripts $A$ and $E$ denote the azimuth and elevation directions of different AoDs \cite{DFTCB}.
Similarly, we define an RX codebook $\text{CB}_{rx}$ composed of $N_{rx} = N_{rx}^A \times N_{rx}^E$ beams designed with different AoAs. 
In general, a TX beam is identified with beam index $i \in \{1,2,..., N_{tx}\}$ and RX beam index $i' \in \{1,2,..., N_{rx}\}$. At a given time instant $t$, the DL received signal at the UE from the connected gNB can be expressed as
\begin{equation}
    y(t) = \sqrt{P} s(t) \textbf{b}_{rx}^* \textbf{H}(t) \textbf{b}_{tx} + \textbf{b}_{rx}^* \textbf{z}(t),
\end{equation}
where $(.)^*$ denotes the conjugate transpose of a vector, $P$ is the total power of the gNB, $s(t)$ represents the transmitted signal such that $\mathop{\mathbb{E}}[|s(t)|^2]=1$, $\textbf{H}(t) \in \mathbb{C}^{M_{rx}\times M_{tx}}$ is the time-varying gain/phase response between TX and RX for an orthogonal frequency-division multiplexing (OFDM) system \cite{6824736}, where we have neglected the dependency on subcarrier for simplicity, $\textbf{z}(t) \in \mathbb{C}^{M_{rx}}$ is the additive white
Gaussian noise with zero mean and variance $\sigma_z^2$, $\textbf{b}_{tx}$ and $\textbf{b}_{rx}$ are the TX and RX beamforming vectors adopted by the gNB and UE, respectively, while UE measures the RSRP across the bandwidth occupied by RSs, which can be expressed as $r(t)=|y(t)|^2$. To obtain the quality of DL TX beams in $\text{CB}_{tx}$, we define the vector of RSRP measurements associated with all the $N_{tx}$ RS resources as $\textbf{R} = [r_{1}(t), r_{2}(t),..., r_{N_{tx}}(t)]$. In the following subsections, we consider RSRP measurements for a set of TX beams denoted as Set B. For spatial-domain beam prediction (SBP), we use the vector of RSRP values measured at the time instant $t$ expressed as $\textbf{R}_{Set B}$, whereas, for time-domian beam prediction (TBP), we use the history of RSRP measurements, including past time instants, e.g., $t-L,...,t-1$, and expressed with vectors of RSRPs as $\textbf{R}_{Set B}^{t-L},..., \textbf{R}_{Set B}^{t-1},\textbf{R}_{Set B}^{t}$. 

\subsection{Spatial-domain Beam Prediction}
In SBP, the UE measures a first set of TX beams defined as Set B which are included in a codebook $\text{CB}_{tx, B}$ 
 of dimension $N_{tx, B}$ and predicts a second set of Tx beams defined as Set A included in a codebook $\text{CB}_{tx, A}$ 
 of dimension $N_{tx, A}$, where Set B has a lower dimension than Set A, i.e., $N_{tx, B} \leqslant N_{tx, A}$. In the network, this is realized by the gNB transmitting a group of DL RS resources associated with Set B beams while the UE performs $N_{tx, B}$ measurements to determine the vector of RSRPs $\textbf{R}_{Set B}$ to be used as input for the AI/ML model. At each time instant $t$, the AI/ML model predicts K beams from Set A, given the measurements of Set B beams used as input. The best K, i.e., Top-K beams indices predicted by the AI/ML model can be expressed as
\begin{equation}
\label{eqn:SBP}
 \big\{\hat{\textit{i}}_1, \hat{\textit{i}}_2,..., \hat{\textit{i}}_K\big\}
     = \mathop{\arg \max}_{i \in \{1,2,...,N_{tx, A}\}}\big(f_W(\textbf{R}_{Set B})\big), 
\end{equation}
where the prediction function $f_W(.)$ depends on the AI/ML model training parameters $W$ and is used to predict the Top-1 beam in Set A, i.e. the beam having the highest RSRP in Set A. The operator $\mathop{\arg} \mathop{\max}(.)$ selects the Top-K beam indices having the highest probability to be the Top-1 beam in Set A, where the index $\hat{\textit{i}}_1$ identifies the beam having the highest probability, while the $\hat{\textit{i}}_K$ identifies the one having the $\textit{K}$-th highest probability. 
Next, we present the two alternatives of SBP studied in 3GPP: Alt. i) Set B is different from Set A and Alt. ii) Set B is a subset of Set A \cite{3gpp38843}.     

\subsubsection{Set B is Different to Set A}
\begin{figure}[h!]
\centerline{\includegraphics[width=\myFigScale\linewidth]{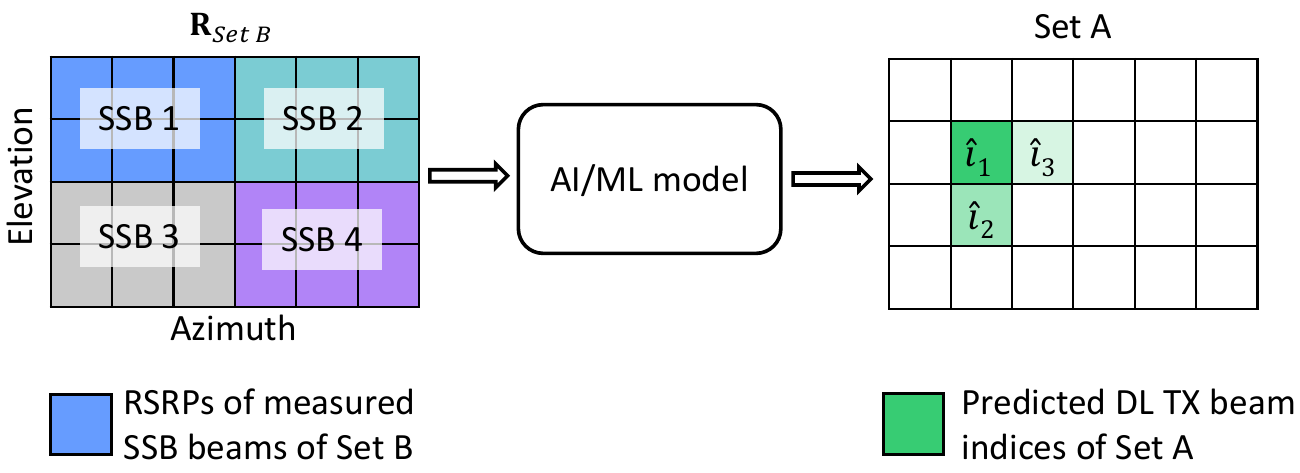}}
\caption{Top-K TX beam prediction in SBP where  Set B is different to Set A.}
\label{fig:SBP1}
\end{figure}

In this alternative, Set B beams may have different characteristics like beam shape, width and set of angular directions in comparison to the beams in Set A. For instance, we consider that Set B are wide SSB beams generally used for transmitting synchronization signals and providing coverage whereas Set A are refined CSI-RS beams used for improving the SINR at the UE. To note that a wide SSB beam in Set B may be generated by the linear combination of adjacent CSI-RS beams in Set A as shown in Fig \ref{fig:SBP1} where one SSB beam encompasses three adjacent CSI-RS beams in azimuth direction and two CSI-RS beams in elevation direction \cite{CBDesign}. Thus, given the input constituted by the measurements of SSB beams, the AI/ML model provides as output the Top-K predicted CSI-RS beam indices.

\subsubsection{Set B is a Subset of Set A}

\begin{figure}[h!]
\centerline{\includegraphics[width=\myFigScale\linewidth]{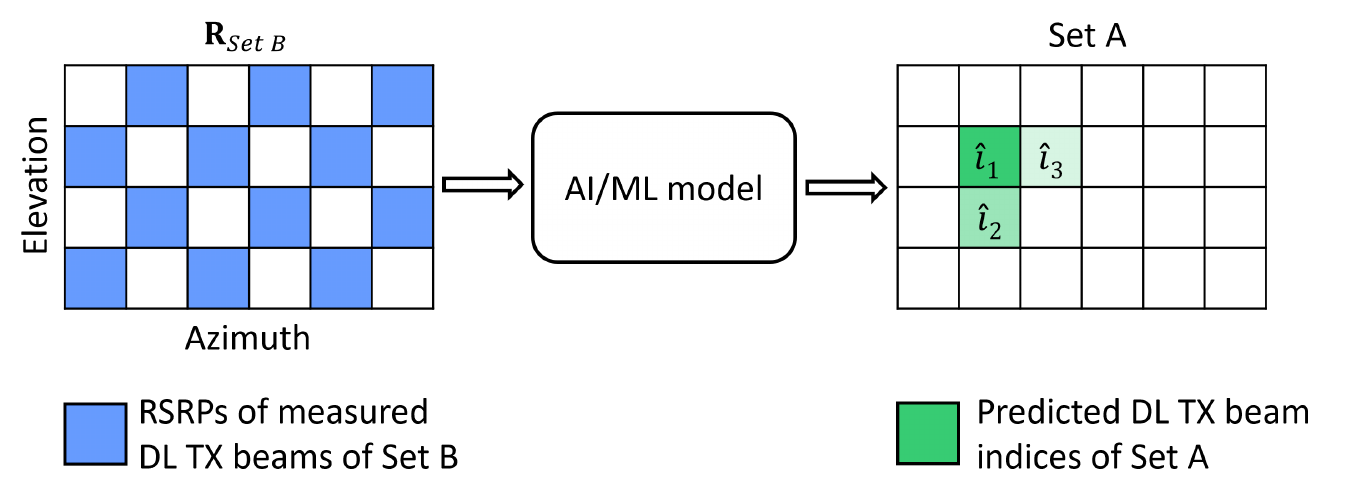}}
\caption{Top-K TX beam prediction in SBP where  Set B is a subset of Set A.}
\label{fig:SBP2}
\end{figure}

In this other alternative shown in Fig \ref{fig:SBP2}, Set B is a subset of the Set A beams. For instance, we consider that Set B are part of the refined CSI-RS beams which partially cover the full set of refined CSI-RS beams belonging to Set A. Therefore, the gNB transmits CSI-RS resources corresponding to Set B beams, while the UE uses the measurements from CSI-RS beams for input to the AI/ML model to obtain the Top-K predicted CSI-RS beam indices like for the previous alternative with Set B different from Set A. The selection criteria of Set B beams from Set A determines the performance of the AI/ML model. Thus, in this work, we define a CSI-RS beam codebook for Set A with a fixed Set B selection pattern such that Set B covers different azimuth and elevation angle directions of Set A beams. 

\subsection{Time-domain Beam Prediction}
\begin{figure}[h!]
\centerline{\includegraphics[width=\myFigScale\linewidth]{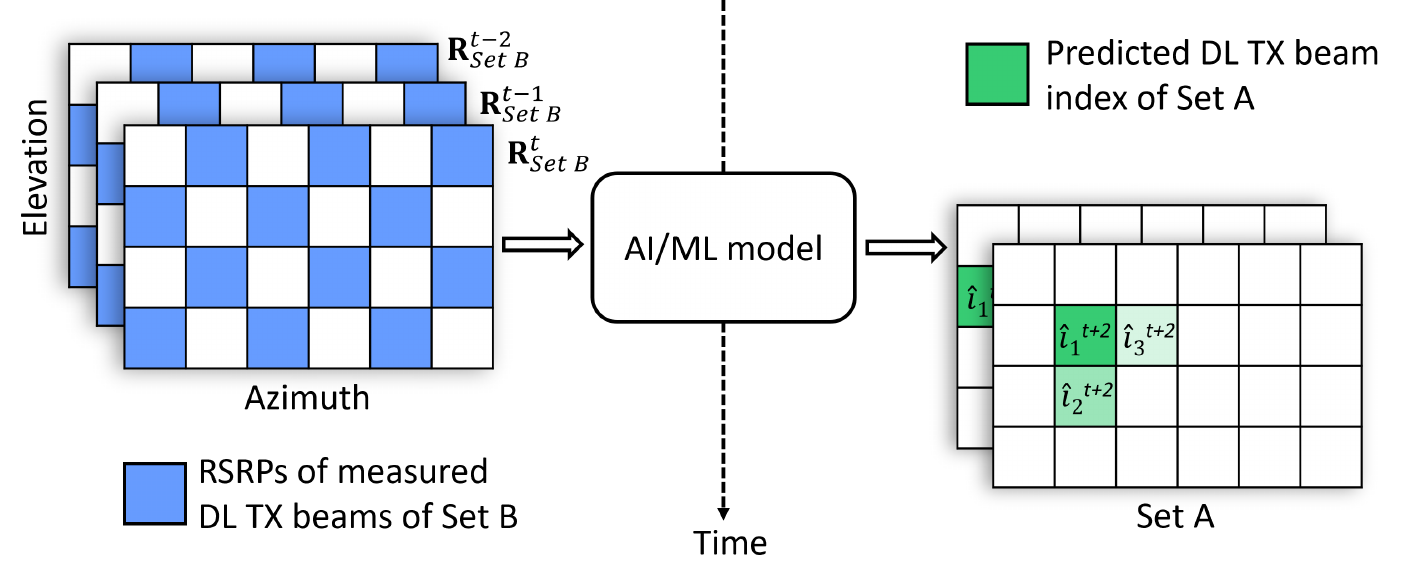}}
\caption{Top-K TX beam prediction in TBP where Set B is a subset of Set A.}
\label{fig:TBP}
\end{figure}

In TBP shown in Fig. \ref{fig:TBP},  the AI/ML model considers as input historical RSRP measurements of Set B and predicts the Top-K strongest beams in Set A to be used at future time instances. 3GPP studied different alternatives like in SBP, however, our focus will be limited to the cases where Set B and Set A are both formed by refined CSI-RS beams and Set B is included or the same as Set A. 
We consider as the AI/ML model input the history of RSRP measurements of the CSI-RS beams within an observation window that includes $l_{o}$ CSI-RS measurements. The AI/ML model provides the Top-K predicted CSI-RS beam indices for one or more future time instances within a prediction window of dimension $l_{p}$. The Top-1 TX beam predictions of the AI/ML model can be expressed as
\begin{equation}
\small
\label{eqn:TBP}
    \big\{\hat{\textit{i}}^{t+1}_1,..., \hat{\textit{i}}^{t+l_p}_1\big\} =\mathop{\arg \max}_{i \in \{1,2,...,N_{tx, A}\}} \big(g_W(\{\textbf{R}_{Set B}^{t-l_o+1},..., \textbf{R}_{Set B}^{t}\})\big) 
\end{equation}
where the prediction function $g_W(.)$ depends on the AI/ML model parameters $W$, $\textbf{R}_{Set B}^{t}$ denotes the measured RSRPs of Set B at time $t$. Eq. (\ref{eqn:TBP}) defines the Top-1 predicted beam. Yet similar to SBP, the Top-K predicted beams for time $t+x$ can be expressed as $[\hat{\textit{i}}^{t+x}_1, \hat{\textit{i}}^{t+x}_2,..., \hat{\textit{i}}^{t+x}_K]$ where the index $\hat{\textit{i}}^{t+x}_1$ identifies the beam having the highest probability, while the $\hat{\textit{i}}^{t+x}_K$ identifies the one having the $\textit{K}$-th highest probability at time $t+x$. 

\section{AI/ML BM Framework} 
\begin{figure}[h]\centering
\subfloat[NW-sided AI/ML model]{\label{fig:NEWsided}\includegraphics[width=.5\linewidth]{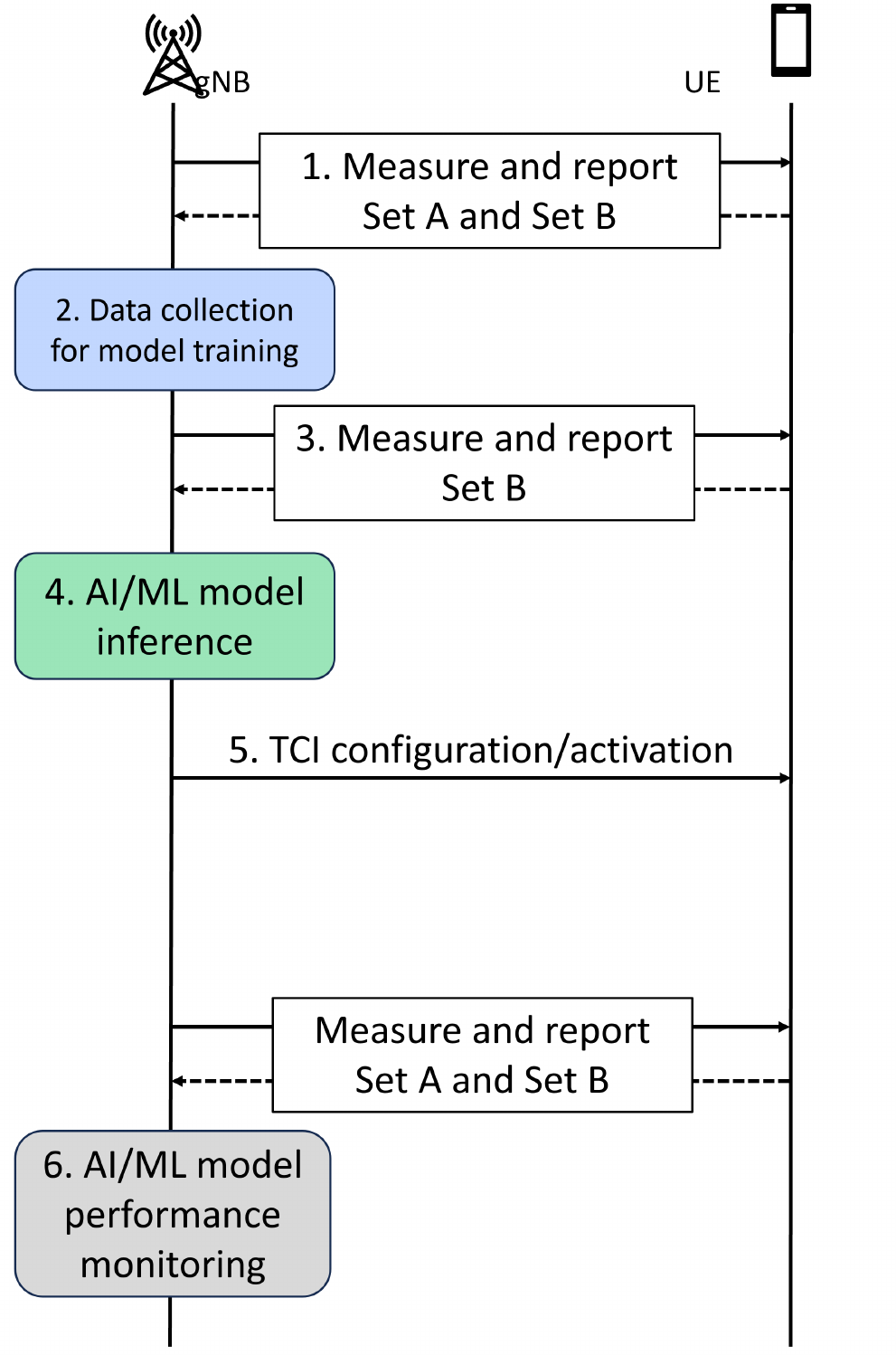}}\hfill
\subfloat[UE-sided AI/ML model]{\label{fig:UEsided}\includegraphics[width=.5\linewidth]{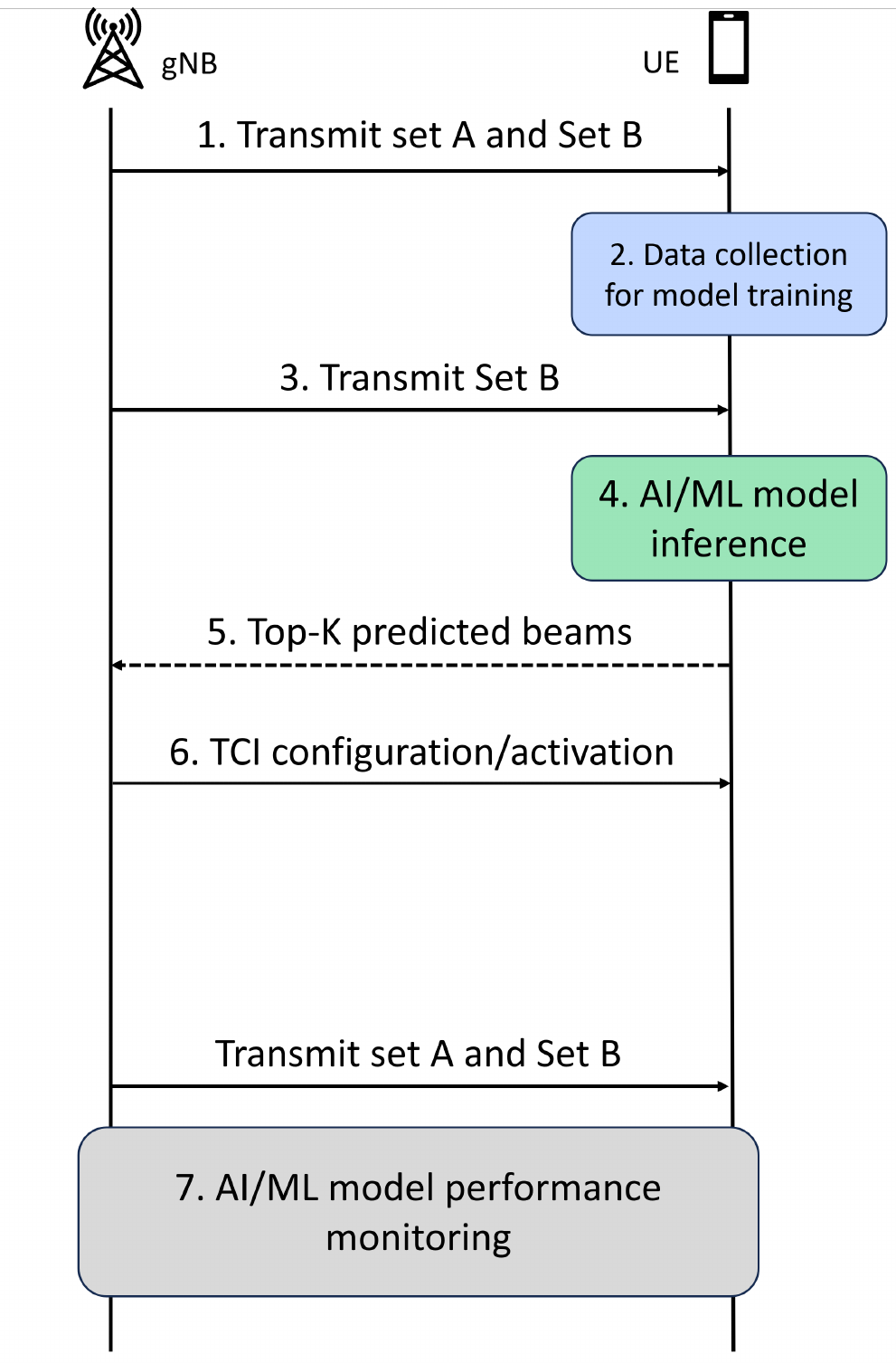}}
\caption{AI/ML BM procedures.}
\label{fig:AIMLBMproce}
\end{figure}

In this section, we discuss the AI/ML model operations for SBP and TBP AI/ML models, which can be trained and deployed at the gNB (NW-sided models) or UE (UE-sided models) as depicted in Fig. \ref{fig:AIMLBMproce}. Moreover, 3GPP Rel-18 SI considered aspects related to the inference, performance monitoring and data collection, which are described in clause 7.1.3 of technical report (TR) 38.843. As the models discussed in SBP and TBP require UE measurements and reporting configurations to support AI/ML models in different stages, most of the aspects discussed in 3GPP focus on introducing enhancements of reporting configuration, measurement RS resource set configuration and overhead reduction of beams reporting\cite{3gpp38843}.

\subsection{Data Collection for AI/ML Model Training}
\label{SecIIIA}

Data collection for model training may reuse the existing beam measurement and reporting frameworks. For instance, the gNB configures the UE with a CSI reporting configuration to perform measurements of Set A and Set B beams as shown in step 1 of Fig. \ref{fig:NEWsided} and Fig. \ref{fig:UEsided}. In step 2 of Fig. \ref{fig:NEWsided} and Fig. \ref{fig:UEsided}, the measurements are collected across various UEs and time occasions may be uploaded to a server for offline model training \cite{3gpp38843}. 
In both SBP and TBP, the UE measures RS resources corresponding to different TX beams using an optimal RX beam determined by previous measurements. Also, the UE may filter the instantaneous RSRP measurements to mitigate the effects of fast fading with a Layer-1 (L1) filter and collect L1-RSRP measurements.
The following differences may depend on whether UE-sided or NW-sided models are used.  

For a UE-sided model, data is collected by measuring the RSs corresponding to Set A and Set B beams, where Set A measurements to derive the best beam index to be used as the label and L1-RSRP measurements associated with Set B beams to be used as input data. 
For the NW-sided model, the UE is configured to measure Set A and Set B beams and report the best beam index in Set A beams to be used as the label in addition to the L1-RSRP measurements of Set B beams for the input data. 

5G NR currently supports beam reporting with a maximum of four beams in a reporting instance \cite{3gpp38802}, yet for the NW-sided model, AI/ML-based beam prediction requires enhancing the reporting to potentially support the reporting of all Set A beam measurements in one reporting instance. In this regard, the reporting overhead for NW-sided model training may be significantly increased compared to the UE-sided model training. On the other hand, the network (NW) has a better awareness of the conditions such as beam patterns and NW antenna configuration associated with the collected data compared to UE.

\subsection{AI/ML Model Inference}

Once the trained AI/ML model is deployed at UE or NW, the gNB configures the UE to measure Set B beams to be used as input to the model during inference. As shown in steps 3-4 in Fig. \ref{fig:UEsided} for UE-sided AI/ML model, the UE predicts Top-K beams in Set A based on L1-RSRP measurements associated with Set B and then reports the predicted Top-K beams to the gNB. For the NW-sided AI/ML model, as illustrated in steps 3-5 in Fig. \ref{fig:NEWsided}, the UE is configured to report all $N_{tx, B}$ of Set B beam measurements to be used as input of the AI/ML model for inference. Similar enhancements to beam measurements reporting described in Sec. \ref{SecIIIA} for the data collection may also be required for NW-sided model inference.

After the UE performs inference and reports the predicted Top-K beams to gNB or after the gNB obtains the predicted Top-K beams based on inference, the gNB may optionally perform an additional step of refined measurements of the Top-K predicted beams. Alternatively, the predicted Top-K beam can be used directly without a step of refined measurements to make informed decisions about transmission control information (TCI) state activation and indication to the UE. At step 5 of Fig. \ref{fig:NEWsided} (step 6 in \ref{fig:UEsided}), a DL control message including the TCI state is transmitted to indicate the DL TX beam to use for receiving DL channels. 

\subsection{AI/ML Performance Monitoring}
For various reasons such as sudden changes in channel conditions or other changes to the NW or UE configurations, both the UE-sided and the NW-sided models may not always provide accurate predictions. Performance monitoring is used to detect any degradation in the AI/ML model's performance during such events \cite{3gpp38843}. Performance monitoring may be performed at the NW side or UE-side. Alternatively may be assisted by the UE. All options are applicable for SBP and TBP. In the following, we provide an example of performance monitoring. 

For NW-side performance monitoring of a UE model shown in step 6 of Fig. \ref{fig:UEsided}, the gNB configures an RS resource set to be monitoring RS resources, where UE measures these monitoring RS resources and reports back to the gNB the measurements corresponding to Set A. 
After receiving the measurements and beam prediction reports, the gNB can calculate relevant KPIs, e.g., beam prediction accuracy, RSRP difference between RSRP measured and RSRP predicted, and other KPIs discussed in Sec. \ref{KPIs}. Moreover, for UE-sided performance monitoring, UE calculate relevant KPI and report to gNB to make a decision. 

For NW-side performance monitoring with beam prediction at NW shown in step 7 of Fig. \ref{fig:NEWsided}, the gNB determines the relevant KPIs based on the measurements reported by the UE and the beam predictions obtained from the NW-side model. 

After evaluation of the performance metrics, an AI/ML model may continue to be used if its performance is deemed acceptable, otherwise beam operations may be switched back to the legacy non-ML BM procedure. Control signalling may be used for AI/ML model switching (including fallback to legacy) based on performance monitoring results.

\section{Simulation Framework} 
In this section, we discuss the system-level simulation methodology and assumptions based on 3GPP 5G-Advanced NR Rel-18. We explain the AI/ML model training and we go through the steps of AI/ML model integration. Then, we list the AI/ML model evaluation KPIs and the system-level performance KPIs considered by Rel-18 SI to evaluate the performance of the AI/ML models. In the following our focus is mainly on NW-sided models ($N_S = N_{tx, A}$), however, similar observations can be made when evaluating the UE-sided models although the signaling configuration for measurement and reporting is less obvious and will be specified in Rel-19 work item (WI).

\subsection{System-Level Simulation Assumptions}
We use a proprietary simulator calibrated to the 3GPP reference scenarios and already used in the past for other 3GPP-related studies \cite{FREAC}. We configure the simulator with the 3GPP 5G-Advanced NR Rel-18 assumptions with an Urban Macro scenario and adopt the following workflow: At first, we use the extracted simulation statistics for generating synthetic data for training the AI/ML models offline. Later we integrate the AI/ML models in the simulation loop, and we simulate a second campaign for obtaining the AI/ML model performance KPIs and system-level KPIs. Table \ref{table:simparas} summarizes the 3GPP-based SLS assumptions considered for the data collection and performance evaluation. 
In addition to the common SLS assumptions stated in Table \ref{table:simparas}, SBP focus on quasi-static UEs with a speed of $3$ kmph, whereas TBP consider a UE speed of $30$ kmph. Moreover, we use a 64-beam CSI-RS codebook and an 8-beam SSB codebook for SBP studies whereas a 32-beam CSI-RS codebook is used for TBP evaluations.          

\subsection{AI/ML Models and Training}

\begin{table*}[t]
\small
\caption{Summary of the AI/ML models used for the different use cases}
\label{table:Stedefs}
\begin{tabular}{|P{3cm}|P{2.3cm}|P{1.9cm}|P{4cm}|P{4.7cm}|}
\hline
\textbf{Model Name} &
\textbf{Alternative} &
  \textbf{Model Type} &
  \textbf{Model Input} &
  \textbf{Model Output} \\ \hline
SBP1$\_N_{tx, B}\_N_{tx, A}$ BM-Case1&
  Set B is different to Set A &
  DNN &
  L1-RSRP of $N_{tx, B}$ beams and beam indices for $N_{tx, B} = 16$ &
  Probability of $N_{tx, A}$ beams being Top-1 beam for $N_{tx, A} = 64$ \\ \hline
SBP2 $\_N_{tx, B}\_N_{tx, A}$ BM-Case1&
  Set B is subset of Set A &
  CNN-DNN &
  L1-RSRP of $N_{tx, B}$ beams and beam indices for $N_{tx, B} = \{8, 16, 32\}$ &
  Probability of $N_{tx, A}$ beams being Top-1 beam for $N_{tx, A} = 64$ \\ \hline
TBP$\_N_{tx, B}\_N_{tx, A}$ BM-Case2&
  Set B is same/subset of Set A&
  LSTM-CNN &
  Historic L1-RSRP of $N_{tx, B}$ beams for $N_{tx, B} = \{8, 16, 32\}$ &
  Probability of $N_{tx, A}$ beams being Top-1 beam for a future time instance when $N_{tx, A} = 32$ \\ \hline 
\end{tabular}
\end{table*}

We propose to use an AI/ML model based on neural network (NN) to predict the index of the best beam based on the objective functions stated in Eq. (\ref{eqn:SBP}) and Eq. (\ref{eqn:TBP}). A summary of AI/ML models used for both SBP and TBP is presented in Table \ref{table:Stedefs}. For SBP with Set B is different from Set A, we consider a deep neural network (DNN) architecture as more complex architectures have not shown significant advantages when considering a limited set of SSB beams as input. 
On the other hand, for SBP with Set B is a subset of Set A, we adopt an AI/ML model with a CNN-DNN architecture that shows better results when considering a larger set of CSI-RS beams as input. For the TBP use case, we adopt a long short-term memory (LSTM) based on CNN architecture to better leverage the time dependencies of measurements in the historical data used as input of the model. The values considered for the observation window and prediction window are $l_o = 5$ and $l_p = 1$, respectively. Finally, the complexity of the model is later provided in Sec. \ref{Chap:designGuidelines}.

The dataset for model training is obtained from 200 network drops with 42000 UEs in total. We split the data between training, validation and testing using the ratio $0.8:0.1:0.1$. We use the ``Adam" model training optimizer and we consider the ``StepLR'' as the learning rate scheduler set with an initial learning rate of $0.01$. For training, we adopt a categorical cross-entropy loss function which can be expressed as $\pazocal{L} =-\sum_n^N\sum_k ^{N_{tx}} y_{n,k}\log(p_{n,k})$, where $N$ is the number of training samples, $y_{n,k}$ is the ground-truth for the $n$-th sample, $p_{n,k}$ is the probability to predict the ground-truth for the $n$-th sample, given by a softmax function. 

\subsection{Integration of AI/ML Model in SLS}
\begin{figure*}
\centerline{\includegraphics[width=\myFigScale\linewidth]{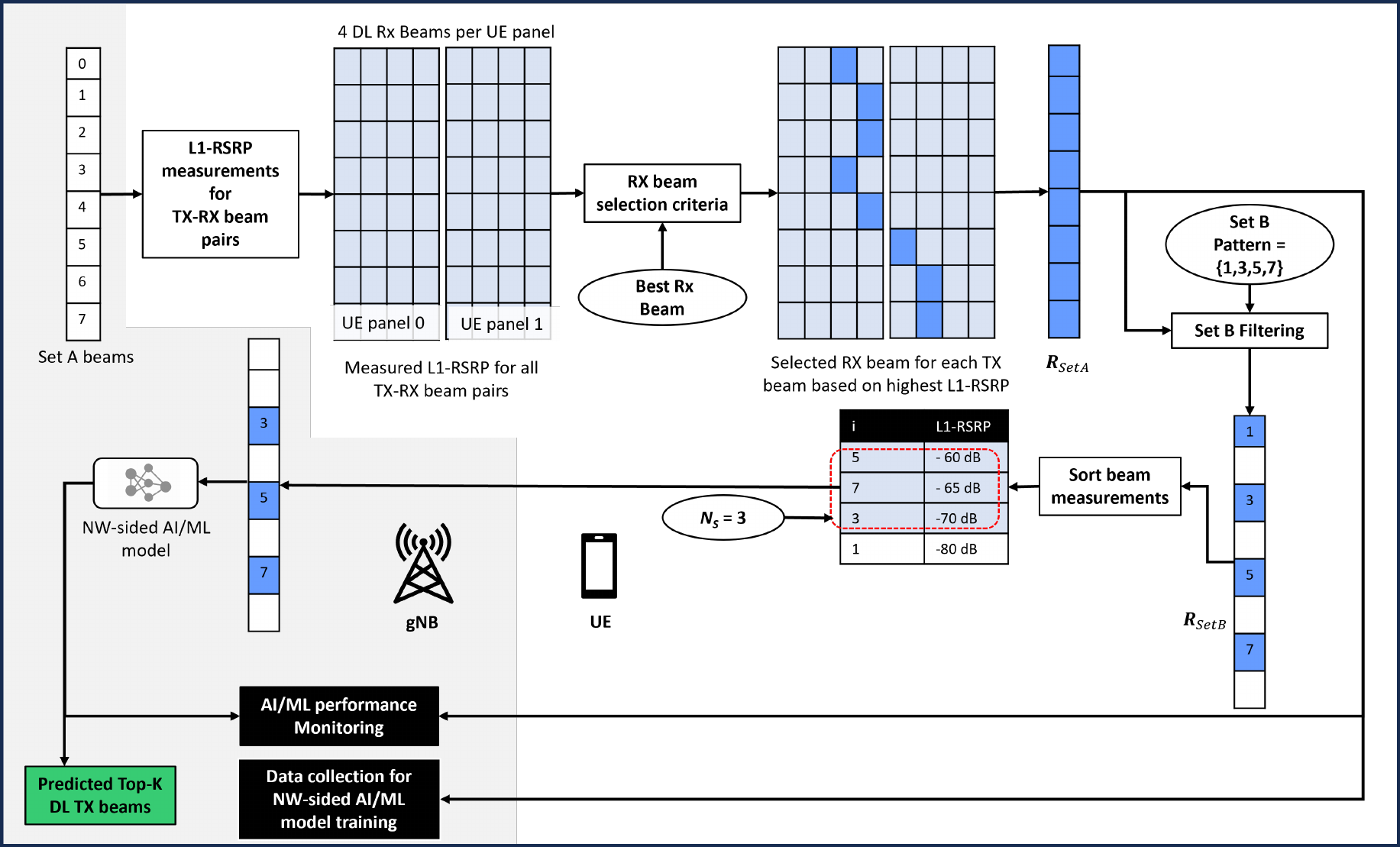}}
\caption{Illustration of beam measurement and reporting framework for a NW-sided model in SBP where Set B is a subset of Set A. }
\label{fig:imp}
\end{figure*}

Fig. \ref{fig:imp} illustrates the beam measurement and reporting framework developed in the SLS where the grey area represents NW-side functionalities and the white area presents UE-side functionalities. Several features have been introduced to support the data collection for AI/ML model training, model inference, and model performance monitoring. 
Starting from the top of Fig. \ref{fig:imp}, we describe the different steps related to the NW-sided models as follows:
\begin{itemize}
    \item First, to generate the dataset needed for AI/ML model training at the NW-side, gNB transmit Set A and UE measures L1-RSRP of each TX beam with different RX beams for both UE panels. 
    The UE determines the L1-RSRP measurement for each TX beam based on the RX beam selection criteria, e.g. considering the ``best RX" beam which provides the highest L1-RSRP assumed in this study. 
    The TX beams with the corresponding L1-RSRP measurements are reported to the gNB that performs data collection. 
    \item Next, for the AI/ML model inference with Set B is a subset of Set A case, the UE refines instantaneous RSRP measurements corresponding to Set B by applying an averaging filter which takes Set B pattern as input. If Set B is different to Set A, the UE measures Set A as well as Set B at different time instances. For both cases, the resulting $\textbf{R}_{Set B}$ is sorted based on filtered L1-RSRP measurements, and the $N_s$ strongest beams and corresponding L1-RSRP measurements are reported to the gNB.
    \item During the AI/ML model inference, the AI/ML model takes the measurement reports and predicts the Top-K TX beam indices. 
    An additional step is considered for the TBP AI/ML model inference where the measurement reports received at the gNB during the observation window are stored for the duration $l_o$ in a buffer before using them for inference. Also, the predicted Top-K TX beam indices from a TBP AI/ML model are stored in memory for $l_p$ time instances as the prediction is valid for $t+l_p$ future time instances. 
    Based on the Top-1 predicted valid for time $t$, the gNB indicates to the UE which TX beam to use for DL data transmission. 
    \item Finally, for monitoring the NW-sided model performance, the same measurements determined in step 1 are used. 
    To note that in case Set A is different from Set B, the gNB transmits, and the UE measures and reports Set B separately, similar to Set A for data collection and performance monitoring functionalities.
\end{itemize}
The same framework can be utilized for integrating a UE-sided AI/ML model. In that case, $\textbf{R}_{Set B}$ is taken as the input to the AI/ML model and UE reports predicted Top-K TX beam indices to a gNB. Considering the ideal report, the performance of the UE-sided AI/ML model is equivalent to the ones obtained for the NW-sided AI/ML models evaluated in this study.

\subsection{Evaluation KPIs}
\label{KPIs}
3GPP agreed to evaluate the performance of AI/ML-based BM with two KPI options that include beam prediction accuracy-related KPIs and system-level related KPIs \cite{3gpp38843}.  

In beam prediction accuracy-related KPIs, the Top-K accuracy evaluates the percentage of predictions that observed the best TX beam in one of the Top-K predicted beams. Thus, the Top-K accuracy can be expressed as
\begin{equation}
    A_{\text{Top-K}} =  \sum_{k = 1}^{K} \mathbbm{1} (\hat{\textit{\textbf{i}}}_{k} = i),
    \label{Eq:topK}
\end{equation}
where $\hat{\textit{\textbf{i}}}_{k}$ is the index of the K-th beam in Top-K predicted beams, $i$ is the Top-1 genie-aided beam and $\mathbbm{1}(.)$ is the indicator function. If $\hat{\textit{\textbf{i}}}_{k} = i$, then $\mathbbm{1}(.) = 1$ and $0$ otherwise. Moreover, the Eq. (\ref{Eq:topK}) reduces to Top-1 accuracy when $K=1$ which indicates the best TX beam prediction accuracy. Even though, $A_{\text{Top-K}}$ is low, the predicted TX beam can still be used if it provides an acceptable beam gain. 

The L1-RSRP difference between the Top-1 predicted beam and the Top-1 genie-aided beam provides insight into the beam gain of the prediction which can be expressed as
\begin{equation}
    e_{RSRP} =\textbf{R}_{i} - \textbf{R}_{\hat{i}},
\end{equation}
where $\textbf{R}_{i}$ denotes the L1-RSRP of the Top-1 genie-aided beam at one instant. Similarly, $\textbf{R}_{\hat{i}}$ represents L1-RSRP of predicted Top-1 TX beam. Furthermore, the prediction of the AI/ML model is reliable if $e_{RSRP}$ is less than $1$~dB \cite{3gpp38843}. Hence, the beam prediction accuracy with $1$~dB margin for the error is defined as
    $A_{\text{Top-1, 1dB margin}} = \mathbbm{1} ((\textbf{R}_{i} -\textbf{R}_{\hat{i}}) < 1~\text{dB})$,
that defines the accuracy of the predicted best beam having an L1-RSRP error less than $1$~dB. In this work, we evaluate the AI/ML model accuracy using the above-mentioned KPIs which can be used as a model performance monitoring KPI in the model inference stage. 

In addition to beam prediction accuracy-related KPIs, system-level related KPIs are needed to verify the gains of the AI/ML-based beam predictions in contrast to legacy BM procedures. Average throughput and cell-edge UE throughput indicate whether the AI/ML-based beam predictions can be used in the network. 
Furthermore, the key motivation of the AI/ML-based BM, i.e. the RS measurement overhead reduction (MOR) can be formulated according to the following conditions \cite{3gpp38843}
\begin{itemize}
    \item $\text{MOR} = 1 - \frac{N_{tx, B}}{N_{tx, A}}$ for SBP.\\
     \item $\text{MOR} = 1 - \frac{l_o N_{tx, B}}{(l_o + l_p) N_{tx, A}}$ for TBP.
\end{itemize}

\section{Results and Performance Evaluation} 
In this section, we evaluate the beam prediction accuracy-related KPIs and the system-level performance-related KPIs for the AI/ML models trained for SBP and TBP. For performance evaluation, we use ten simulation drops that are not used for AI/ML model training and we compare the results with respect to baseline cases provided in each subsection. 

\subsection{Evaluation of Different Alternatives in SBP}

\begin{figure}[t]
\centerline{\includegraphics[width=\myplotScale\linewidth]{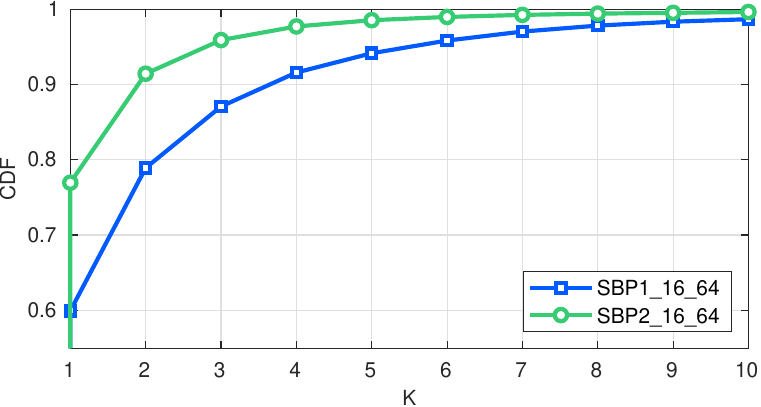}}
\caption{Evaluation of Top-K prediction accuracy ($A_{\text{Top-K}}$) for varying K.}
\label{fig:SBP1_TopK}
\end{figure}

This subsection evaluates a wide-to-narrow prediction model (denoted by SBP1) with respect to a narrow-to-narrow prediction model (denoted by SBP2), where SBP1 and SBP2 models consider SSB beam codebook of size 16 and a refined CSI-RS beam codebook of size 16 as Set B, respectively. Both models consider a refined CSI-RS codebook of size 64 as Set A. Fig. \ref{fig:SBP1_TopK} shows the Top-K prediction accuracy of wide-to-narrow prediction, which is evaluated compared to narrow-to-narrow prediction. Results show that the narrow-to-narrow prediction outperforms the wide-to-narrow prediction. Additionally, the Top-1 prediction accuracy of wide-to-narrow prediction degrades $17\%$ compared to narrow-to-narrow prediction.

\begin{figure}[t]
\centerline{\includegraphics[width=\myplotScale\linewidth]{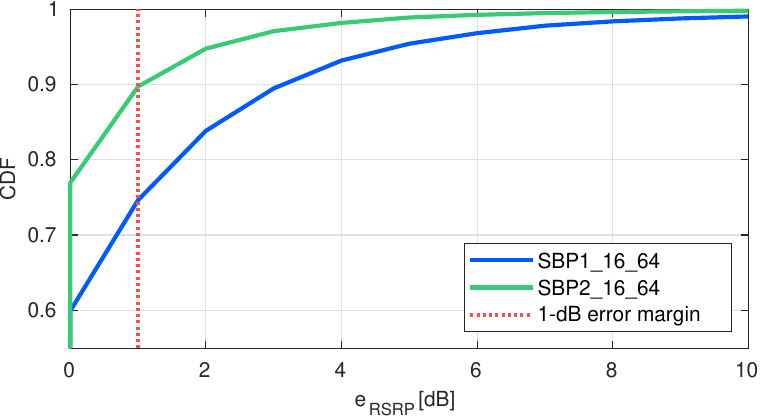}}
\caption{Evaluation of RSRP difference between predicted and genie-aided best beam’s.}
\label{fig:SBP1_RSRPError}
\end{figure}
\begin{figure}[t]
\centerline{\includegraphics[width=\myplotScale\linewidth]{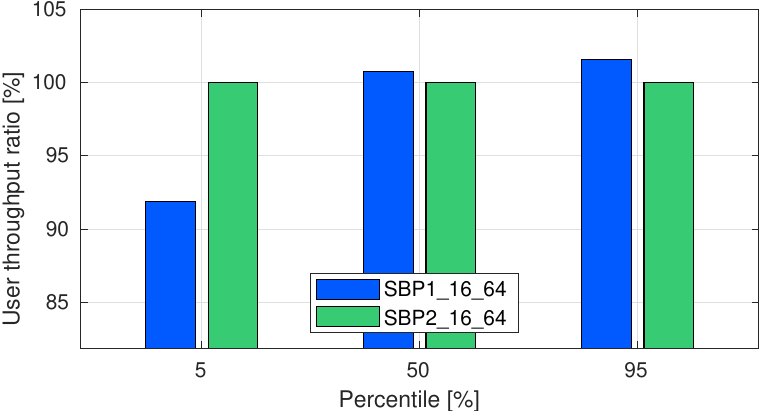}}
\caption{Evaluation of UE throughput ratio of $5$-th, $50$-th and $95$-th percentiles with respect to absolute UE throughput of $[0.52, 1.59, 3.13]\times 10^4$ Mbps of narrow-to-narrow beam prediction, respectively. }
\label{fig:SBP1_UserTput}
\end{figure}

Fig. \ref{fig:SBP1_RSRPError} illustrates the performance evaluation of the models on L1-RSRP difference between Top-1 genie-aided and Top-1 predicted beam's. As it can be seen, both wide-to-narrow prediction and narrow-to-narrow prediction achieve more than $90\%$ accuracy within $e_{RSRP} < 3$ dB. However, wide-to-narrow prediction degrades approximately $15\%$ the Top-1 prediction accuracy with 1 dB margin of the narrow-to-narrow prediction, which achieves $90\%$.

The results depicted in Fig. \ref{fig:SBP1_TopK} and  Fig. \ref{fig:SBP1_RSRPError} show that the prediction of the strongest refined CSI-RS beam is more difficult considering SSB beam measurements as input than using a subset of the CSI-RS codebook measurements as input to the model. This is because SSB beams cover a wide area encompassing multiple CSI-RS beams. Thus, the variation of input L1-RSRP distribution of SSB measurements is less relevant for the AI/ML model-based narrow beam prediction compared to the L1-RSRP distribution of the CSI-RS measurements.
  
Fig. \ref{fig:SBP1_UserTput} analyzes the behaviour of the UE throughput at the 5-th, 50-th, and 95-th percentile of the CDF. The ratio is calculated with respect to narrow-to-narrow prediction throughput performance. For the 5-th percentile of the CDF, cell-edge UEs experience $8.2\%$ UE throughput degradation with wide-to-narrow prediction compared to cell-edge UEs who used narrow-to-narrow prediction. The UE throughput of wide-to-narrow prediction matches the performance of narrow-to-narrow prediction for all other UEs except cell-edge UEs. 

Moreover, the wide-to-narrow prediction model eliminates the P2 procedure as the model requires L1-RSRP measurements obtained from SSB beams. Thus, measurement overhead reduction in P2 is $100\%$. Additionally, the performance of wide-to-narrow prediction model seems reasonable and can be used if it meets the accuracy levels required by the NW. However, the comparison of beam prediction accuracy-related KPIs and UE throughput performance shows that narrow-to-narrow prediction outperforms wide-to-narrow prediction. Therefore in the following subsection, we evaluate different varieties of narrow-to-narrow prediction models by varying the Set B size. 

\subsection{Evaluation of SBP: Set B is a Subset of Set A}
\label{SecVB}
The performance of the narrow-to-narrow beam prediction is evaluated with three AI/ML models with input Set B sizes of 8, 16, and 32 for an output Set A size of 64. The performance of each AI/ML model is compared with a non-AI/ML baseline which selects the strongest beam among the reported Set B beams, which also belongs to Set A (denoted by BL\_$N_{tx, B}$\_$N_{tx, A}$).     

\begin{figure}[t]
\centerline{\includegraphics[width=\myplotScale\linewidth]{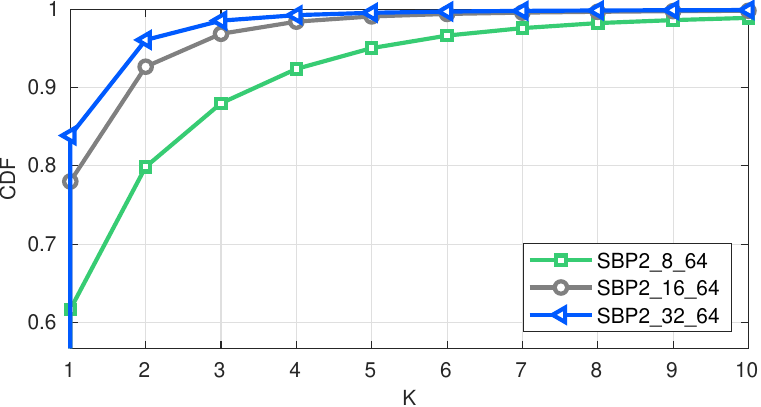}}
\caption{Evaluation of Top-K prediction accuracy ($A_{\text{Top-K}}$) for varying K.}
\label{fig:SBP2_TopK}
\end{figure}

Fig. \ref{fig:SBP2_TopK} compares the Top-K prediction accuracy of AI/ML models for K	$\leqslant$ 10 and Top-K prediction accuracy reduces as the Set B size reduces for any K. Narrow-to-narrow prediction with $N_{tx, B} = 8$ experiences a significant degradation in terms of Top-K prediction accuracy for lower K values compared to narrow-to-narrow prediction with $N_{tx, B} = 16$. However, narrow-to-narrow prediction with $N_{tx, B} = 16$ and $N_{tx, B} = 32$ achieved a Top-1 prediction accuracy above $75\%$ and all three AI/ML models achieved $A_{\text{Top-4}}$ greater than $90\%$. Results confirm that as $N_{tx, B}/N_{tx, A}$ reduces, degradation in Top-K prediction accuracy increases significantly.     

\begin{figure}[t]
\centerline{\includegraphics[width=\myplotScale\linewidth]{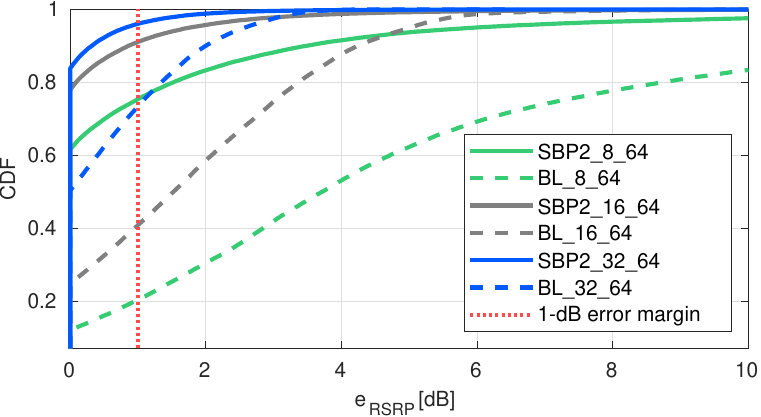}}
\caption{Evaluation of RSRP difference between predicted and genie-aided best beam’s.}
\label{fig:SBP2_RSRPError}
\end{figure}

According to Fig. \ref{fig:SBP2_RSRPError}, the Top-1 prediction accuracy with 1 dB margin of narrow-to-narrow predictions are $\{0.95, 0.90, 0.75\}$ for $N_{tx, B}=\{32,16,8\}$, respectively.
In comparison to corresponding baselines, all AI/ML models in SBP where Set B is a subset of Set A, significantly reduce the RSRP difference. Additionally, results emphasize that the capability of the AI/ML model to predict the correct beam increases as $N_{tx, B}/N_{tx, A}$ approaches $1$, confirming that having more beam measurements is beneficial to yield a better prediction accuracy.

\begin{figure}[t]
\centerline{\includegraphics[width=\myplotScale\linewidth]{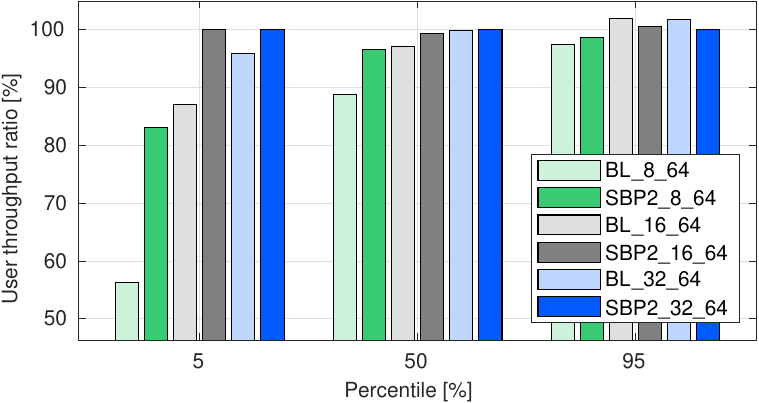}}
\caption{Evaluation of $5$-th, $50$-th and $95$-th percentiles in user throughput ratio with respect to absolute user throughput of $[0.54, 1.64, 3.15]\times 10^4$ Mbps of narrow-to-narrow beam prediction.}
\label{fig:SBP2_UserTput}
\end{figure}

Fig. \ref{fig:SBP2_UserTput} shows the gain in the UE throughput ratio in comparison to its corresponding baseline case. The cell-edge UEs experience a UE throughput gain of $\{27\%,13\%,4\%\}$ with narrow-to-narrow predictions for $N_{tx, B} = \{8, 16, 32\}$ compared to corresponding baseline cases. Observations show that the UE throughput performance increases with narrow-to-narrow predictions in contrast to the baseline case when $N_{tx, B}/N_{tx, A}$ decreases. A similar trend can be seen for cell-median UEs. Comparing narrow-to-narrow prediction models, the UE throughput ratio observed at $95$-th percentile of the CDF does not reflect a performance degradation as $N_{tx, B}$ varies.

UE throughput performance at cell-edge is degraded in narrow-to-narrow predictions when the number of $N_{tx, B}$ measurements decrease.  Similarly, the throughput of cell-edge UEs reduces as UE predicts a beam with a lower gain compared to the best beam. On the other hand, UEs in the $95$-th percentile of the CDF do not experience a throughput loss even if the predicted beam is not the best beam because UE may have more than one strong beam. Therefore, the modulation and coding schemes (MCS) used in the data transmission do not significantly change. 

The MOR in SBP where Set B is a subset of Set A, is equal to $87.5\%$, $75\%$ and $50\%$ for narrow-to-narrow predictions with $N_{tx, B} = \{8, 16, 32\}$, respectively. Similarly, corresponding baseline algorithms achieve the same MOR. According to the above observations, cell-edge UEs are more susceptible to throughput degradation as Set B size reduces to obtain a higher MOR. Additionally, the results emphasize the importance of having different AI/ML models on the UE-side or NW-side. For example, if a UE using narrow-to-narrow prediction with $N_{tx, B} = 8$ starts to experience poor channel conditions, UE may enable with an AI/ML model applying with $N_{tx, B} = 16$ or $N_{tx, B} = 32$ to continue the narrow-to-narrow prediction avoiding the user throughput degradation.

Even though the deployed narrow-to-narrow prediction AI/ML models exhibit performance enhancement compared to non-AI/ML baseline schemes, the NW configurations used for the data collection phase can differ in the inference phase. Therefore, the following subsection explores the model generalization aspect of narrow-to-narrow prediction models as the antenna configuration of the NW changes.

\subsection{Evaluation of Generalization Aspects of SBP}
In this subsection, we evaluate the AI/ML model generalization aspect of SBP. Evaluation is done with the narrow-to-narrow prediction with $N_{tx, B} = 16$ and $N_{tx, A} = 64$  model, which is trained in a gNB composed of an antenna array equipped with $M^V_{tx} = 4, M^H_{tx} = 8$ denoted by 4x8. Then, the AI/ML model is integrated and tested with two more wireless networks composed of gNB with antenna array 4x4 and 4x16.

\begin{figure}[t]
\centerline{\includegraphics[width=\myplotScale\linewidth]{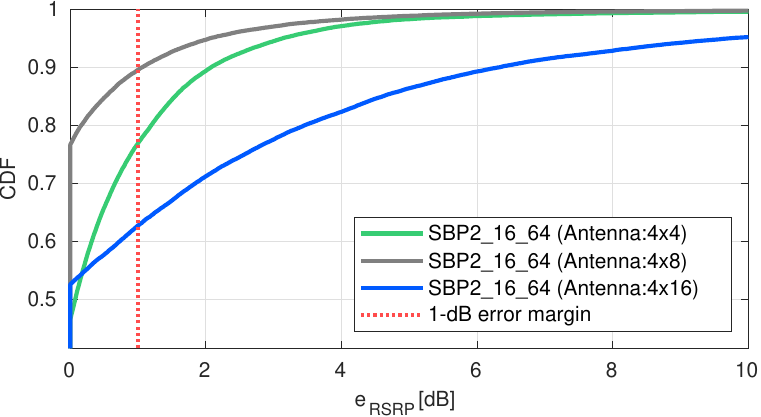}}
\caption{Evaluation of RSRP difference between predicted and genie-aided best beam’s.}
\label{fig:SBPGen_RSRPError}
\end{figure}

Fig. \ref{fig:SBPGen_RSRPError} depicts the variation of RSRP difference of narrow-to-narrow prediction with different antenna configurations. Note that the Top-1 prediction accuracy is equal to the ratio at which $e_{RSRP} = 0$ and it is observed that Top-1 prediction accuracy drops by a minimum $25\%$ if the model is used in a gNB with different antenna configurations. However, the narrow-to-narrow prediction with antenna 4x4 achieves $77\%$ of Top-1 prediction accuracy with 1 dB margin and shows a better generalization ability compared to narrow-to-narrow prediction with antenna 4x16 which yields a $63\%$ Top-1 prediction accuracy with 1 dB margin. The observation concludes that the RSRP difference of more narrow beams generated by the antenna 4x16 is higher compared to the RSRP difference of relatively wider beams created by antenna 4x4 if the predicted Top-1 beam differs from the best genie-aided beam.      

\begin{figure}[t]
\centerline{\includegraphics[width=\myplotScale\linewidth]{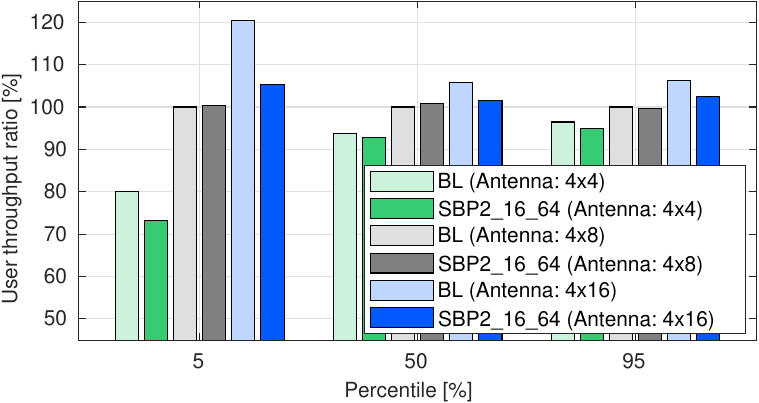}}
\caption{Evaluation of UE throughput ratio with respect to the baseline case of trained antenna configuration 4x8 which yields a $5$-th, $50$-th and $95$-th percentiles of absolute UE throughput of $[0.49, 1.56, 3.20]\times 10^4$ Mbps. }
\label{fig:SBPGen_Tput}
\end{figure}

The narrower beams generated by a 4x16 antenna array provide a higher beam gain compared to a 4x8 antenna array. Similarly, antenna 4x4 reduces the beam gain compared to antenna 4x8. Thus, Fig. \ref{fig:SBPGen_Tput} presents the evaluation of the UE throughput ratio corresponding to the baseline case (denoted by BL in Fig. \ref{fig:SBPGen_Tput}) with the same antenna array configuration. BL cases represent the non-AI/ML beam selection approach where the best beam selected with respect to exhaustive beam search, i.e. Set B is same as Set A. The gNB equipped with the 4x16 antenna enhances throughput of cell-edge UEs by $20.5\%$ compared to gNB with antenna 4x8. Yet, the UE throughput gain achieved with a larger antenna compared to 4x8 is reduced by $15\%$ due to AI/ML model generalization impact. The same trend is visible for UEs at the $50$-th and $95$-th percentile of UE throughput. Moreover, using a small 4x4 array reduces the overall throughput as shown by the baseline, and the poor performance is further degraded due to AI/ML model generalization impacts. 


These results suggest that the AI/ML model performs the best when the specified antenna configuration is the same for both model training and inference. Otherwise, beam prediction accuracy and UE throughput degrade due to model generalization impacts.  Alternatively, the model can be trained with mixed dataset that include multiple gNB antenna configurations and achieve satisfactory performance \cite{3gpp38843}. Still, the problem persist if a new configuration is adopted by gNB. Note that the UEs in different channel conditions such as cell-edge, cell-center, etc. may use different models to increase or decrease the prediction accuracy as shown in Sec. \ref{SecVB}. Yet, the NW antenna panel may impact all UEs, and the cell-edge UE performance degradation is unavoidable, thus these UEs may only fallback to legacy non-AI/ML BM procedures. 


\subsection{Evaluation of TBP}
In this subsection, we evaluate the beam prediction accuracy and UE throughput performance of both alternatives in TBP, i.e. Set B is equal/subset of Set A. AI/ML models discussed in this subsection consider refined CSI-RS codebooks with sizes of 8, 16, and 32 as the Set B. Moreover, the input to each AI/ML model consists of Set B beam measurements over an observation window size of 5. All models consider a refined CSI-RS beam codebook of size 32 as Set A, which is used after a prediction window size of 1 (denoted by TBP$\_N_{tx, B}\_32$). Additionally, we use a non-AI/ML baseline method;  sample and hold (SnH), which selects the strongest Set B beam at time $t$ and is used at $t$ and $t+1$ as the predicted TX beam.

\begin{figure}[t]
\centerline{\includegraphics[width=\myplotScale\linewidth]{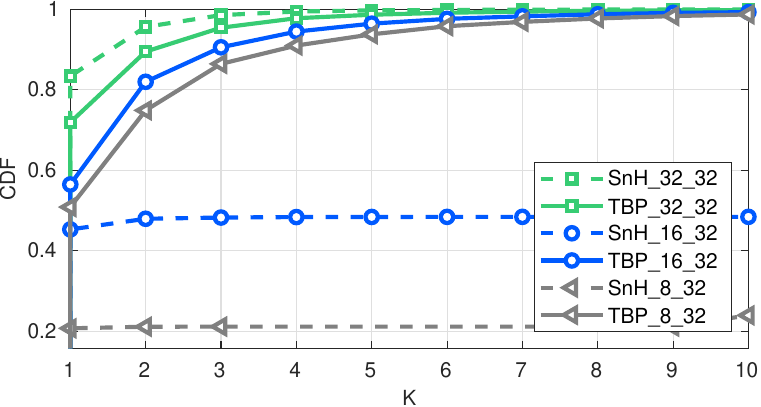}}
\caption{Evaluation of Top-K prediction accuracy ($A_{\text{Top-K}}$) for varying K.}
\label{fig:TBP1_TopK}
\end{figure}

Fig. \ref{fig:TBP1_TopK} depicts the Top-K prediction accuracy evaluation for TBP models along with the corresponding SnH cases. It is observed that the model with $N_{tx, B} = 32$ degrades on Top-K prediction accuracy compared to the corresponding SnH scheme due to the dominant temporal correlation of the channel. However, models with Set B is a subset of Set A; $N_{tx, B}$ = $\{16, 8\}$ achieve $11\%$ and $30\%$ gain on Top-1 prediction accuracy compared to corresponding SnH schemes, respectively. Hence, results show that when $N_{tx, B}/N_{tx, A}$ reduces, the Top-K prediction accuracy improves in contrast to corresponding SnH schemes. However, employing a higher number of measured beams helps to improve the beam prediction accuracy. 

\begin{figure}[t]
\centerline{\includegraphics[width=\myplotScale\linewidth]{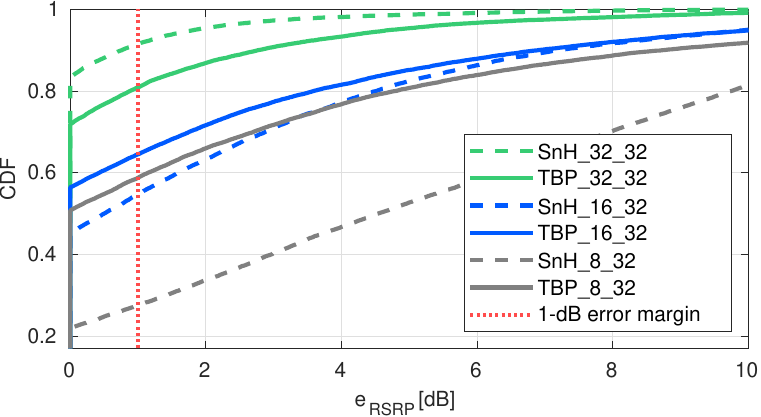}}
\caption{Evaluation of RSRP difference between predicted and genie-aided best beam’s.}
\label{fig:TBP1_RSRPError}
\end{figure}

The RSRP difference evaluation of TBP models along with corresponding SnH schemes are depicted in Fig. \ref{fig:TBP1_RSRPError}. The AI/ML model with Set B codebook size of 32 suffers a loss of $10.4\%$ with respect to Top-1 prediction accuracy with 1 dB margin loss of $91.5\%$ obtained by the corresponding SnH scheme. The result also confirms that models with $N_{tx, B}$ = $\{16, 8\}$
perform better compared to corresponding SnH schemes. Yet, Top-1 prediction accuracy with 1 dB margin for models with $N_{tx, B}$ = $\{16, 8\}$
are $65\%$ and $59\%$, respectively. Additionally, $20\%$ or more UEs experience more than $3$ dB RSRP difference between predicted and genie-aided best beam’s for both AI/ML models with $N_{tx, B}$ = $\{16, 8\}$.

\begin{figure}[t]
\centerline{\includegraphics[width=\myplotScale\linewidth]{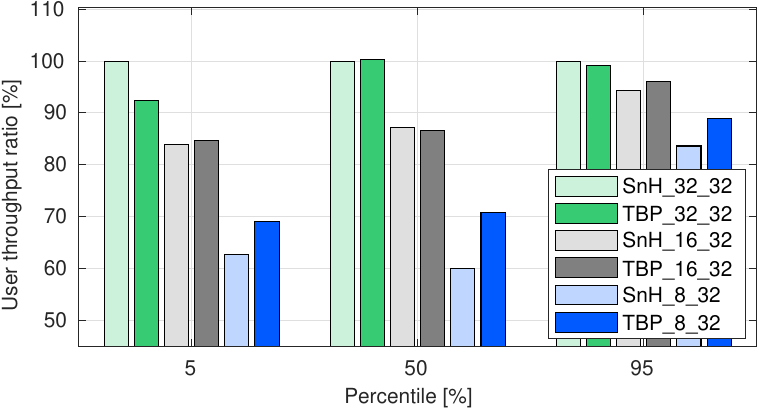}}
\caption{Evaluation of UE throughput ratio with respect to SnH$\_32\_32$ which yields an absolute per UE throughput of $[0.09, 0.70, 2.49]\times 10^4$ Mbps at $5$-th, $50$-th and $95$-th percentiles.}
\label{fig:TBP1_Tput}
\end{figure}

The observed trends in beam prediction KPIs are further confirmed by the UE throughput performance shown in Fig. \ref{fig:TBP1_Tput}. The cell-edge UEs using the AI/ML model with $N_{tx, B} = 32$
experience a $7.7\%$ loss of throughput in contrast to the corresponding SnH scheme. Yet, cell-edge UEs using the AI/ML model with $N_{tx, B} = 8$
experience a $6.4\%$ gain in throughput compared to corresponding SnH. The throughput of cell-median UEs who use models with $N_{tx, B}$ = $\{16, 32\}$
show similar performance compared to corresponding baseline cases. Yet, 50-th percentile UEs in the CDF who use the AI/ML model with $N_{tx, B} = 16$
experience a throughput improvement of $10\%$ in comparison to similar UEs who use the corresponding SnH scheme. The  AI/ML model with $N_{tx, B} = 8$
performs better at $95$-th percentile of UE throughput compared to $5$-th and $50$-th, and yet it is less than $90\%$ of UE throughput obtained by SnH scheme with Set B equals to Set A. 
Importantly, the evaluated AI/ML models with Set B codebook sizes 32, 16, and 8 provide a MOR of $16.7\%$, $58.3\%$, and $80\%$, respectively. 
In a mobile network, the speeds of the UEs may vary from time to time. Hence, the following subsection evaluates the model generalization impact for varying UE speeds.  


\subsection{Evaluation of Generalization Aspects of TBP}
\begin{figure}[t]
\centerline{\includegraphics[width=\myplotScale\linewidth]{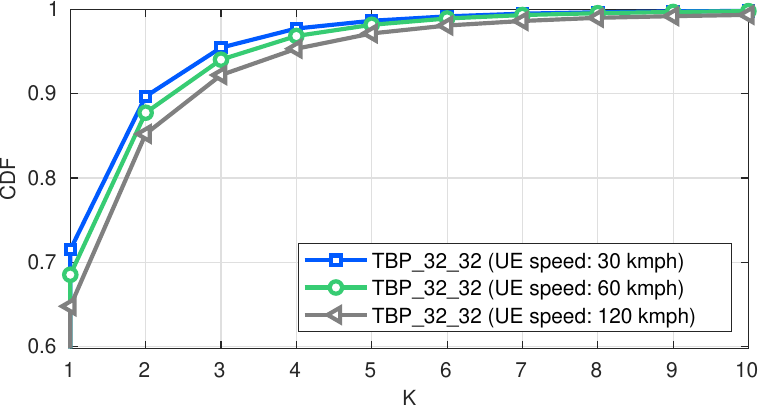}}
\caption{Evaluation of Top-K prediction accuracy for varying K.}
\label{fig:TBPGen_TopK}
\end{figure}

In this subsection, we evaluate the generalization capability of the TBP AI/ML model with Set B and Set A codebooks of size 32 trained with a UE speed of $30$ kmph and used in UEs with higher speeds (60 kmph and 120 kmph). Fig. \ref{fig:TBPGen_TopK} shows the variation of Top-K prediction accuracy with different UE speeds when K $\leqslant$ 10. UEs with $60$ kmph speed experience a loss of $3\%$ in Top-1 prediction accuracy compared to UEs trained at $30$ kmph. For an even higher speed of $120$ kmph,  the loss in Top-1 prediction accuracy increases to $6.7\%$. As the UE speed increases compared to the trained speed, the input distribution of the L1-RSRP measurements changes. Moreover, the beam measurements become outdated within a shorter time duration, since higher speeds introduce a significant change in UE position.

\begin{figure}[t]
\centerline{\includegraphics[width=\myplotScale\linewidth]{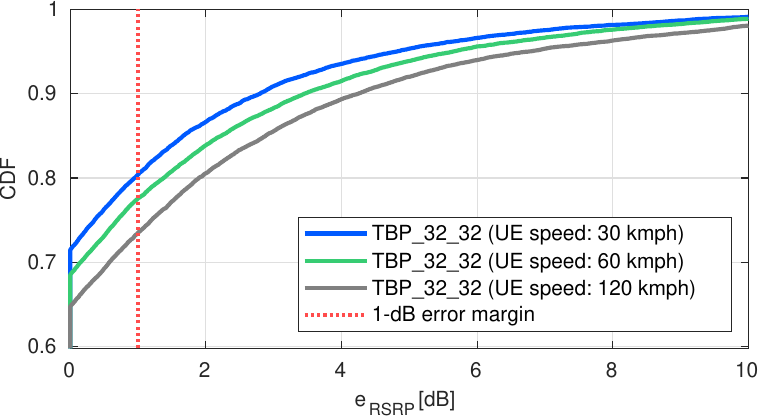}}
\caption{Evaluation of RSRP difference between predicted and genie-aided best beam’s.}
\label{fig:TBPGen_RSRPError}
\end{figure}

Fig. \ref{fig:TBPGen_RSRPError} depicts $2.9\%$ loss in Top-1 prediction accuracy with 1 dB margin when UE speed is $60$ kmph compared to $30$ kmph. The degradation becomes even higher for UEs with $120$ kmph as $7\%$ in comparison to UEs with a speed of $30$ kmph. According to beam prediction accuracy related results, the model generalization impact of TBP models for higher speeds is insignificant. 

\begin{figure}[t]
\centerline{\includegraphics[width=\myplotScale\linewidth]{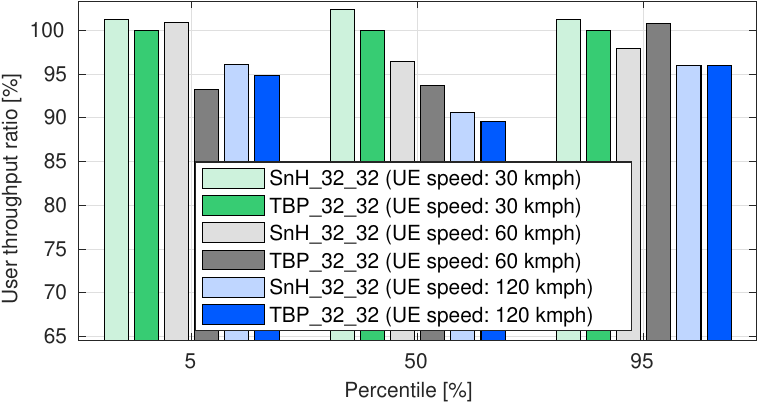}}
\caption{Evaluation of UE throughput ratio with respect to TBP$\_32\_32$ trained for $30$ kmph UE speed, which yields an absolute per UE throughput of $[0.09, 0.69, 2.48]\times 10^4$ Mbps at $5$-th, $50$-th and $95$-th percentiles.}
\label{fig:TBPGen_Tput}
\end{figure}

In Fig. \ref{fig:TBPGen_Tput}, throughput at cell-median UEs use AI/ML model with speeds of $60$ and $120$ kmph show $6.3\%$ and $10.4\%$ degradation compared to UEs maintaining the model trained speed of $30$ kmph. However, throughput at cell-median UEs who used the SnH scheme with higher speeds of $60$ and $120$ kmph also show a similar degradation as seen with the AI/ML model performance. Hence, the cell-median UEs who used the AI/ML model at higher speeds compared to the trained speed experienced a throughput degradation mainly due to the more challenging channel conditions in mobility. The cell-edge UEs experience a very low throughput for higher speeds compared to cell-median UEs due to mobility and channel uncertainties. Moreover, the UEs at $95$-th percentile of throughput are resilient to throughput degradation and achieve more than $95\%$ of throughput with higher speeds compared to the trained UE speed of $30$ kmph. User throughput results explain the AI/ML model generalization impact is insignificant for the TBP model where Set B is the same as Set B.

\subsection{Design Guidelines for AI/ML-Based BM}
\label{Chap:designGuidelines}
In this subsection, we provide some recommendations for AI/ML model usage based on the evaluation results in the previous subsections. We assume that an AI/ML model can be recommended for use in a situation if it can achieve a minimum of $95\%$ UE throughput compared to the best baseline case observed. The AI/ML model use guidelines for different UE scenarios are presented in Table \ref{table:DesignGuidelines}.


\begin{table}[h]
\small
\caption{Recommendations for AI/ML model selection.}
\label{table:DesignGuidelines}
\begin{tabular}{|p{1.8cm}|p{1cm}| p{1.3cm}|p{1.7cm}|p{0.7cm}|}
\hline
\textbf{AI/ML model} & \textbf{MOR (\%)} & \textbf{Cell-edge UEs} & \textbf{Cell-median UEs} & \textbf{ Best UEs} \\
\hline
SBP1\_$16$\_$64$ & 100 &\xmark  & \checkmark & \checkmark \\
\hline
SBP2\_$8$\_$64$ & 87.5 &\xmark  & \checkmark & \checkmark \\
\hline
SBP2\_$16$\_$64$ & 75 &\checkmark  & \checkmark & \checkmark \\
\hline
SBP2\_$32$\_$64$ &  50 &\checkmark  & \checkmark & \checkmark \\
\hline
TBP\_$8$\_$32$ &  80 &\xmark  & \xmark & \xmark \\
\hline
TBP\_$16$\_$32$ & 58.3 &\xmark  &  \xmark  &  \checkmark  \\
\hline
TBP\_$32$\_$32$ & 16.7 &\xmark   &  \checkmark   &  \checkmark  \\
\hline
\end{tabular}
\end{table}

\begin{figure}[h]\centering
\subfloat[Distribution of minimum K value such that Top-K prediction accuracy is 100\%.]{\label{fig:TopKMap}\includegraphics[width=.48\linewidth]{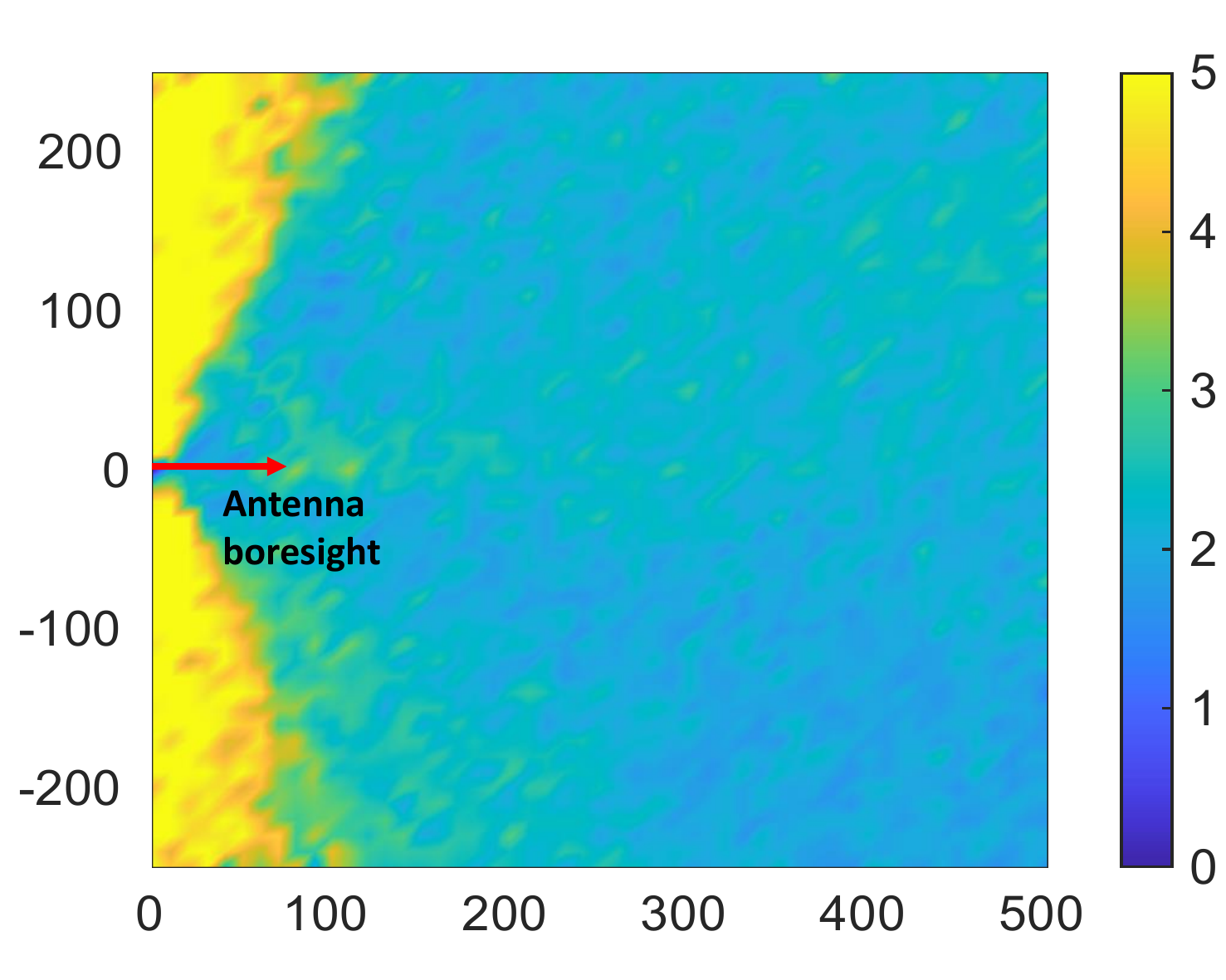}}\hfill
\subfloat[Distribution of RSRP error between predicted beam and Top-1 genie aided beam's.]{\label{fig:RSRPEmap}\includegraphics[width=.48\linewidth]{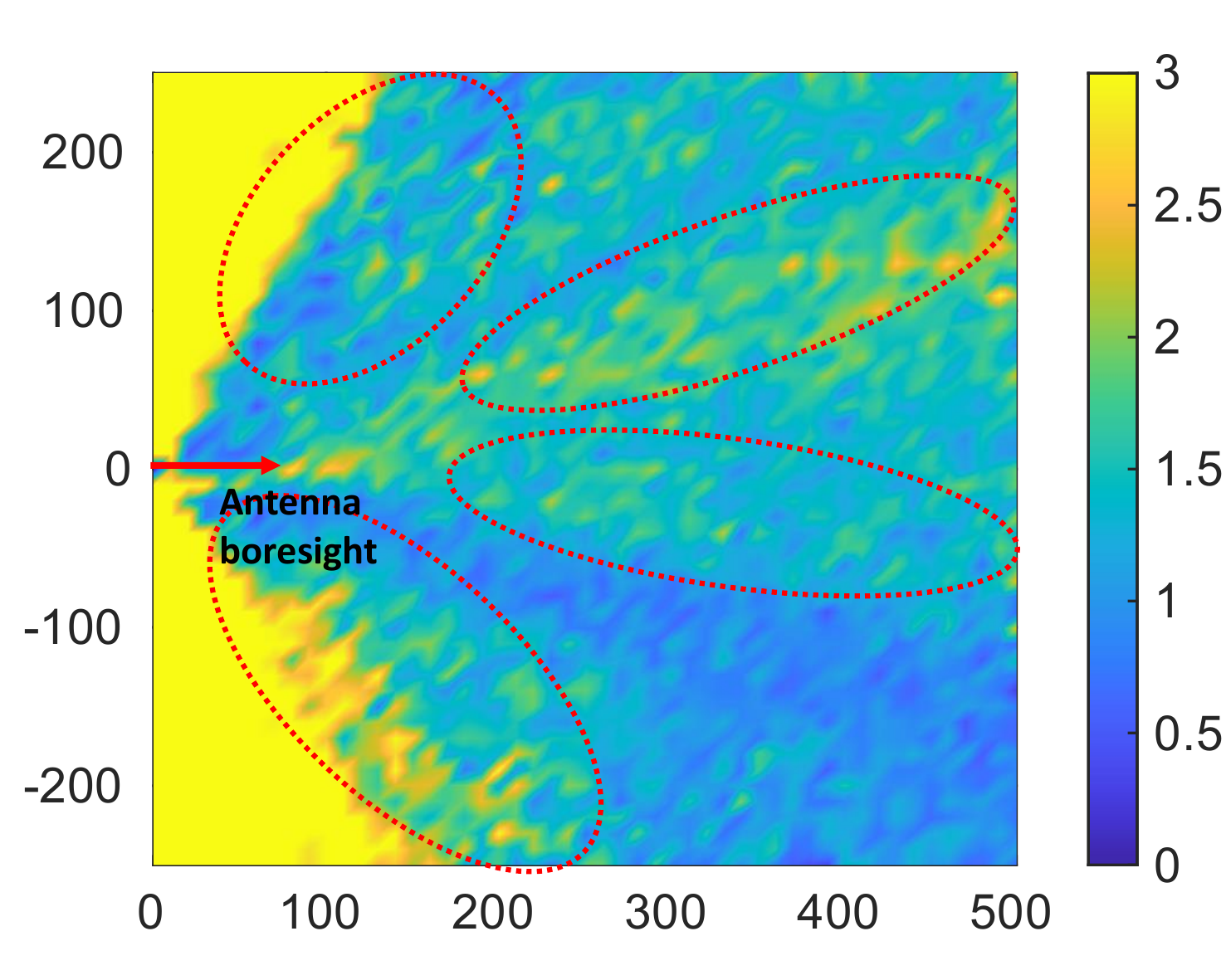}}
\caption{The spatial distribution of beam prediction accuracy of narrow-to-narrow prediction with Set B size 8, with a gNB located in $x = 0$, $y = 0$.}
\label{fig:Maps}
\end{figure}

Firstly, the AI/ML models discussed in SBP where Set B is a subset of Set A decreases the UE throughput of cell-edge UEs while reducing the measurement overhead. Hence, cell-edge UEs may required to use narrow-to-narrow prediction models requiring a larger number of measurements compared to other UEs in the cell, which can use narrow-to-narrow prediction models with higher MOR (e.g. MOR = 87.5), as their throughput difference compared to legacy BM is negligible. 
With further experiments using a simpler network deployment, we observed that the Top-K prediction accuracy does not show significant differences among UEs in the coverage area (i.e. the AI/ML model does not discriminate between UEs) as shown in Fig \ref{fig:TopKMap}. However, on the other hand, Fig \ref{fig:RSRPEmap} depicts that the RSRP difference between predicted and Top-1 genie-aided beam increases significantly especially for some elevation/azimuth directions, where the penalized areas could be at the beam-boundaries or cell-boundaries. Such areas may likely include cell-edge UEs, where the throughput is impacted.

At the same time, the wide-to-narrow beam prediction model, which could avoid CSI-RS beam transmissions in the P2 procedure, 
may be applied for all UEs, except for cell-edge UEs, as they experience some throughput degradation compared to the narrow-to-narrow prediction model. 
Moreover, the AI/ML model generalization aspects of SBP show that the expected gain on UE throughput performance using a high number of antenna elements is degraded due to the SBP AI/ML models trained with a lower number of antenna elements. Thus, UE throughput gains provided by enabling the transmission with a larger antenna size may be limited by generalization. This is because the SBP models need to be used for the configurations in which the AI/ML model is trained or in case of a configuration change, the AI/ML model needs to be fine-tuned to support the new configuration. In all of the cases, performance monitoring procedure may allow the UEs to fallback to legacy non-ML BM procedures in case of severe performance degradation of the model. 

Conversely, the TBP studies of narrow-to-narrow beam prediction AI/ML models show marginal performance gains compared to the sample and hold baseline. Thus, a second-round of beam measurements based on AI/ML model prediction may be used to improve the performance. 
Generally, the AI/ML model with Set B and Set A size of 32 can be used for UEs with strong channel conditions and cell-median UEs even though the advantages of measurements overhead are less compared to SBP models. 
The AI/ML model with Set B size of 16 and Set A size of 32 can be used for UEs with strong channel conditions, and the model with Set B size of 8 and Set A size of 32 is not suitable for any UE due to its low performance. 
Finally, according to the given results, the degradation in generalization due to UE speed is less compared to the changes in the NW antenna configuration which applies also to TBP. Overall, the poor performance with AI/ML models at cell-edge because of wrong beam prediction highlights the need for improving the AI/ML model training. For instance, by penalizing the prediction errors of cell-edge UEs compared to other UEs, which could be realized by adopting more sophisticated loss functions or using different sampling of training data.

An additional important aspect is the computational complexity of AI/ML model. All SBP AI/ML models consist of $50\times 10^3$ to $150\times 10^3$ hyper-parameters where the AI/ML model size is less than 1 MB. Additionally, the computational complexity of models is less than $1.2\times 10^6$ floating point operations per second (FLOPS). TBP AI/ML models composed of $100\times 10^3$ to $200\times 10^3$ hyper-parameters which also require less than 1 MB of memory to store the model and less than $2\times 10^6$ FLOPS. Hence, the AI/ML models evaluated in this study have moderate computational complexity and limited memory occupation. The computational complexity can be further reduced considering the designs as in \cite{khan2024lowcomplexity}, which provides a promising direction to make the AI/ML models studied in this work to be easily transferred, deployed and stored at the UE.

In summary, according to Table \ref{table:DesignGuidelines}, most of the AI/ML models failed to deliver more than $90\%$ throughput for cell-edge UEs compared to baseline cases. Improvement of cell-edge UE performance using assisted information such as UE positions, AoA and AoD can be considered as a future research direction. Moreover, we assumed ideal L1-RSRP measurements and reporting from a UE, inaccurate L1-RSRP measurements may degrade the beam prediction accuracy as well as system-level performance. Also, we assumed UE report all Set B beams and corresponding RSRP measurements to gNB. However, the UE may consider reporting partial beam measurements ($N_S \leqslant N_{tx, B}$) and adopt adaptive quantization to limit the reporting overhead. Finally, the  AI/ML model activation and deactivation based on performance monitoring and coexistence between AI/ML-based BM and legacy BM are other interesting research directions. 

\section{Conclusion} 
In this paper, we presented an in-depth analysis of SBP and TBP AI/ML models and their generalization abilities to investigate the link between ML model performance and system-level performance. The analysis covered sub-use cases of SBP, Set B is different from Set A and Set B is a subset of Set A, and sub-use cases of TBP where Set A equals Set B and Set B is a subset of Set A as discussed in 3GPP Rel-18 SI. We derived AI/ML BM design guidelines based on the throughput performances, which showed the necessity of using AI/ML models with higher Set B size for cell-edge UEs.
Moreover, the AI/ML model generalization issues studied in this paper suggested that the future research directions may address UE-sided AI/ML model generalization capabilities to different conditions (e.g, NW antenna configurations, beam patterns, etc.), reinforcement learning (RL) could be a direction as it may update the model depending on the new-conditions. Moreover, the beam prediction framework studied and currently under consideration for specification in 3GPP Rel-19 WI is intended to be generic therefore RL and other model training enhancements may be supported as well. Model complexity is also another important aspect and can be reduced to improve inference latency.
As beamforming aspects are not impacting only FR2 but also FR1, we envision the usage of beam prediction approaches in different bands. In addition, multiple TX/RX point (multi-TRP) scheme and efficient handover in mobility scenarios can be other venues where to apply beam prediction. Lastly, in future systems prediction reports and measurement reports may be used for power control, link adaptation, among other procedures, which could be designed to use or discard predicted beams measurements depending on AI/ML model accuracy.   

\section{Appendix}
Table \ref{table:simparas} provides the set of simulation parameters used to configure the system-level simulator.
\begin{table}[h]
\small
\caption{3GPP-based SLS parameters}
\label{table:simparas}
\begin{center}
\begin{tabular}{|p{2.8cm}| p{5.2cm}|}
\hline
\textbf{Parameter} & \textbf{Value}\\
\hline
Frequency range & FR2 with a carrier frequency of 30 GHz  \\
\hline
Network deployment & $500$~m inter-site distance (ISD), gNB antenna height: $25$~m, $7$ sites with $3$ sectors/gNB per site with wrap-around
\\
\hline
Channel mode &  Urban macro (UMa) scenario with distance-dependent LoS probability \\
\hline
System bandwidth & $80$ MHz, subcarrier spacings: $120$ kHz\\
\hline
UE distribution & $10$ UEs per sector/gNB, UE antenna height: $1.5$ m, $100\%$ outdoor UEs\\
\hline
gNB antenna configuration & One panel with (M, N, P, Mg, Ng): $(4, 8, 2, 1, 1)$, $(d_V, d_H)$ = $(0.5, 0.5) \lambda$ \\
\hline
gNB beam selection & based on maximum L1-RSRP\\
\hline
gNB antenna radiation pattern & TR 38.802 Table A.2.1-6, Table A.2.1-7 \cite{3gpp38802}\\
\hline
Maximum TX power & $40$ dBm for gNB, $23$ dBm for UE\\

\hline
UE antenna configuration & Panel structure: (M, N, P) = (1, 4, 2), 2 UE panels (left, right) with (Mg, Ng) = (1, 2)\\
\hline
UE panel selection & Ideal\\
\hline
Number of UE beams & 8 (4 beams per UE panel) \\
\hline
UE beam selection & Optimal RX beam selection based on maximum L1-RSRP calculated with sliding window size 3\\
\hline
UE antenna radiation pattern & TR 38.802 Table A.2.1-8, Table A.2.1-10 \cite{3gpp38802} \\
\hline
Link adaptation & Based on CSI-RS \\
\hline
Control and RS overhead & 	common overhead is 30\% based on TR 37.910 for DenseUrban-eMBB scenario \cite{3gpp37910} \\
\hline
Traffic model & 	Full buffer \\
\hline
Control channel decoding & 	Ideal \\
\hline
UE receiver type & 	MMSE-IRC \\
\hline
Beamforming scheme & 	Analog Beamforming \\
\hline
gNB selection & 	Based on the RSRP \\
\hline
RX noise figure & 	7 dB for gNB, 10 dB for UE\\
\hline
\end{tabular}
\end{center}
\end{table}

\Urlmuskip=0mu plus 1mu\relax
\bibliographystyle{IEEEtran}  
\bibliography{Sec_7_Bibliography.bib}

\begin{thebibliography}{10}
\providecommand{\url}[1]{#1}
\csname url@samestyle\endcsname
\providecommand{\newblock}{\relax}
\providecommand{\bibinfo}[2]{#2}
\providecommand{\BIBentrySTDinterwordspacing}{\spaceskip=0pt\relax}
\providecommand{\BIBentryALTinterwordstretchfactor}{4}
\providecommand{\BIBentryALTinterwordspacing}{\spaceskip=\fontdimen2\font plus
\BIBentryALTinterwordstretchfactor\fontdimen3\font minus \fontdimen4\font\relax}
\providecommand{\BIBforeignlanguage}[2]{{%
\expandafter\ifx\csname l@#1\endcsname\relax
\typeout{** WARNING: IEEEtran.bst: No hyphenation pattern has been}%
\typeout{** loaded for the language `#1'. Using the pattern for}%
\typeout{** the default language instead.}%
\else
\language=\csname l@#1\endcsname
\fi
#2}}
\providecommand{\BIBdecl}{\relax}
\BIBdecl

\bibitem{NokiaWhitepaperRel18}
Nokia, ``{5G-Advanced: Expanding 5G for the connected world},'' Nokia, White paper CID210984, February 2023.

\bibitem{NokiaWhitepaperRel19}
{Nokia}, ``{Taking 5G-Advanced to the next level and bridging into the 6G era},'' Nokia, White paper CID213705, January 2024.

\bibitem{8755300}
M.~Chen, U.~Challita, W.~Saad, C.~Yin, and M.~Debbah, ``{Artificial Neural Networks-Based Machine Learning for Wireless Networks: A Tutorial},'' \emph{IEEE Communications Surveys \& Tutorials}, vol.~21, no.~4, pp. 3039--3071, 2019.

\bibitem{9569393}
J.~Pihlajasalo, D.~Korpi, M.~Honkala, J.~M.~J. Huttunen, T.~Riihonen, J.~Talvitie, A.~Brihuega, M.~A. Uusitalo, and M.~Valkama, ``{HybridDeepRx: Deep Learning Receiver for High-EVM Signals},'' in \emph{2021 IEEE 32nd Annual International Symposium on Personal, Indoor and Mobile Radio Communications (PIMRC)}, 2021, pp. 622--627.

\bibitem{9605055}
A.~Kelkar and C.~Dick, ``{NVIDIA Aerial GPU Hosted AI-on-5G},'' in \emph{2021 IEEE 4th 5G World Forum (5GWF)}, 2021, pp. 64--69.

\bibitem{10017176}
J.~Pihlajasalo, D.~Korpi, M.~Honkala, J.~M.~J. Huttunen, T.~Riihonen, J.~Talvitie, A.~Brihuega, M.~A. Uusitalo, and M.~Valkama, ``{Deep Learning OFDM Receivers for Improved Power Efficiency and Coverage},'' \emph{IEEE Transactions on Wireless Communications}, vol.~22, no.~8, pp. 5518--5535, 2023.

\bibitem{9446676}
J.~Hoydis, F.~A. Aoudia, A.~Valcarce, and H.~Viswanathan, ``{Toward a 6G AI-Native Air Interface},'' \emph{IEEE Communications Magazine}, vol.~59, no.~5, pp. 76--81, 2021.

\bibitem{10156818}
M.~Merluzzi \emph{et~al.}, ``{The Hexa-X Project Vision on Artificial Intelligence and Machine Learning-Driven Communication and Computation Co-Design for 6G},'' \emph{IEEE Access}, vol.~11, pp. 65\,620--65\,648, 2023.

\bibitem{9768336}
M.~K. Shehzad, L.~Rose, M.~M. Butt, I.~Z. Kovács, M.~Assaad, and M.~Guizani, ``{Artificial Intelligence for 6G Networks: Technology Advancement and Standardization},'' \emph{IEEE Vehicular Technology Magazine}, vol.~17, no.~3, pp. 16--25, 2022.

\bibitem{3gpp38843}
3GPP, ``{Study on Artificial Intelligence (AI)/Machine Learning (ML) for NR air interface},'' {3rd Generation Partnership Project (3GPP)}, Technical Report (TR) 38.843, 12 2023.

\bibitem{5GNRbook}
M.~Enescu, K.~Jayasinghe, K.~Ranta-Aho, K.~Schober, and A.~Toskala, \emph{{5G Physical Layer}}.\hskip 1em plus 0.5em minus 0.4em\relax John Wiley \& Sons, Ltd, 2020, ch.~6, pp. 87--148.

\bibitem{8734054}
Y.~Wang, A.~Klautau, M.~Ribero, A.~C.~K. Soong, and R.~W. Heath, ``{MmWave Vehicular Beam Selection With Situational Awareness Using Machine Learning},'' \emph{IEEE Access}, vol.~7, pp. 87\,479--87\,493, 2019.

\bibitem{9898910}
P.~Susarla, B.~Gouda, Y.~Deng, M.~Juntti, O.~Silvén, and A.~Tölli, ``{Learning-Based Beam Alignment for Uplink mmWave UAVs},'' \emph{IEEE Transactions on Wireless Communications}, vol.~22, no.~3, pp. 1779--1793, 2023.

\bibitem{9512417}
H.-L. Song and Y.-C. Ko, ``{Beam Alignment for High-Speed UAV via Angle Prediction and Adaptive Beam Coverage},'' \emph{IEEE Transactions on Vehicular Technology}, vol.~70, no.~10, pp. 10\,185--10\,192, 2021.

\bibitem{Li2020}
Y.~N.~R. Li, B.~Gao, X.~Zhang, and K.~Huang, ``{Beam Management in Millimeter-Wave Communications for 5G and beyond},'' \emph{IEEE Access}, vol.~8, pp. 13\,282--13\,293, 2020.

\bibitem{lin2023overview}
X.~Lin, ``{An Overview of the 3GPP Study on Artificial Intelligence for 5G New Radio},'' 2023.

\bibitem{Xue2023}
Q.~Xue, J.~Guo, B.~Zhou, Y.~Xu, Z.~Li, and S.~Ma, ``{AI/ML for Beam Management in 5G-Advanced},'' 2023.

\bibitem{10412253}
J.~Xu, I.~Nakamura, R.~Feng, L.~Liu, and L.~Chen, ``{Performance Evaluation of AI/ML Model to Enhance Beam Management in 5G-Advanced System},'' in \emph{2023 Fourteenth International Conference on Mobile Computing and Ubiquitous Network (ICMU)}, 2023, pp. 1--6.

\bibitem{Zuo2022}
J.~Zuo, J.~Zhang, Y.~Cao, X.~Chen, F.~Wang, N.~Hu, and X.~Xu, ``{Artificial Intelligence-Based Spatial Domain Beam Prediction for 5G Beyond},'' in \emph{2022 IEEE Globecom Workshops (GC Wkshps)}, 2022, pp. 1460--1465.

\bibitem{10123939}
Q.~Li \emph{et~al.}, ``{Machine Learning Based Time Domain Millimeter-Wave Beam Prediction for 5G-Advanced and Beyond: Design, Analysis, and Over-The-Air Experiments},'' \emph{IEEE Journal on Selected Areas in Communications}, vol.~41, no.~6, pp. 1787--1809, 2023.

\bibitem{10334007}
Y.~Bai, J.~Zhang, C.~Sun, L.~Zhao, H.~Li, and X.~Wang, ``{AI-Based Beam Management in 3GPP: Optimizing Data Collection Time Window for Temporal Beam Prediction},'' \emph{IEEE Open Journal of Vehicular Technology}, vol.~5, pp. 48--55, 2024.

\bibitem{10419651}
S.~Wang, W.~Chen, X.~Chen, Y.~Zhang, and B.~Ai, ``{Deep Learning-Based Beam Pair Prediction With Finite Beam Quality Information},'' in \emph{2023 IEEE 23rd International Conference on Communication Technology (ICCT)}, 2023, pp. 588--592.

\bibitem{10200730}
L.~Maggi, A.~R. Koblitz, Q.~Zhu, and M.~Andrews, ``{Tracking the Best Beam for a Mobile User via Bayesian Optimization},'' in \emph{2023 IEEE 97th Vehicular Technology Conference (VTC2023-Spring)}, 2023, pp. 1--7.

\bibitem{10335766}
C.~Sun, L.~Zhao, T.~Cui, H.~Li, Y.~Bai, S.~Wu, and Q.~Tong, ``{AI Model Selection and Monitoring for Beam Management in 5G-Advanced},'' \emph{IEEE Open Journal of the Communications Society}, vol.~5, pp. 38--50, 2024.

\bibitem{9417509}
S.~Khunteta and A.~K.~R. Chavva, ``{Recurrent Neural Network Based Beam Prediction for Millimeter-Wave 5G Systems},'' in \emph{2021 IEEE Wireless Communications and Networking Conference (WCNC)}, 2021, pp. 1--6.

\bibitem{10012751}
D.~Marasinghe, N.~Jayaweera, N.~Rajatheva, S.~Hakola, T.~Koskela, O.~Tervo, J.~Karjalainen, E.~Tiirola, and J.~Hulkkonen, ``{LiDAR aided Wireless Networks - Beam Prediction for 5G},'' in \emph{2022 IEEE 96th Vehicular Technology Conference (VTC2022-Fall)}, 2022, pp. 1--7.

\bibitem{9562975}
A.~Bonfante, L.~G. Giordano, I.~Macaluso, and N.~Marchetti, ``{Performance of Predictive Indoor mmWave Networks With Dynamic Blockers},'' \emph{IEEE Transactions on Cognitive Communications and Networking}, vol.~8, no.~2, pp. 812--827, 2022.

\bibitem{AnalogBF}
V.~V. Ratnam and A.~F. Molisch, ``{Periodic Analog Channel Estimation Aided Beamforming for Massive MIMO Systems},'' \emph{IEEE Transactions on Wireless Communications}, vol.~18, no.~3, pp. 1581--1594, 2019.

\bibitem{DFTCB}
J.~Suh, C.~Kim, W.~Sung, J.~So, and S.~W. Heo, ``{Construction of a Generalized DFT Codebook Using Channel-Adaptive Parameters},'' \emph{IEEE Communications Letters}, vol.~21, no.~1, pp. 196--199, 2017.

\bibitem{6824736}
A.~Kammoun, H.~Khanfir, Z.~Altman, M.~Debbah, and M.~Kamoun, ``{Preliminary Results on 3D Channel Modeling: From Theory to Standardization},'' \emph{IEEE Journal on Selected Areas in Communications}, vol.~32, no.~6, pp. 1219--1229, 2014.

\bibitem{CBDesign}
S.~Noh, M.~D. Zoltowski, and D.~J. Love, ``{Multi-Resolution Codebook and Adaptive Beamforming Sequence Design for Millimeter Wave Beam Alignment},'' \emph{IEEE Transactions on Wireless Communications}, vol.~16, no.~9, pp. 5689--5701, 2017.

\bibitem{3gpp38802}
3GPP, ``{Study on New Radio Access Technology Physical Layer Aspects},'' {3rd Generation Partnership Project (3GPP)}, Technical Report (TR) 38.802, 09 2017.

\bibitem{FREAC}
F.~Abinader, C.~Rom, K.~Pedersen, S.~Hailu, and N.~Kolehmainen, ``{System-Level Analysis of mmWave 5G Systems with Different Multi-Panel Antenna Device Models},'' in \emph{2021 IEEE 93rd Vehicular Technology Conference (VTC2021-Spring)}, 2021, pp. 1--6.

\bibitem{khan2024lowcomplexity}
M.~Q. Khan, A.~Gaber, M.~Parvini, P.~Schulz, and G.~Fettweis, ``{A Low-Complexity Machine Learning Design for mmWave Beam Prediction},'' 2024.

\bibitem{3gpp37910}
{3GPP}, ``{5G; Study on self evaluation towards IMT-2020 submission},'' {3rd Generation Partnership Project (3GPP)}, Technical Report (TR) 37.910, 11 2020.

\end{thebibliography}

\end{document}